\documentclass[12pt]{iopart}
\usepackage{iopams}
\usepackage{epsf}
\usepackage{euscript}

\newcommand{\sign}{\mbox{sign}\,}
\newcommand{\Dbare}{D}
\newcommand{\Cbare}{C}
\newcommand{\onlinecite}{\cite}
\newcommand{\openone}{{\bf 1}}
\newcommand{\hplanck}{2 \pi \hbar}
\newcommand{\appb}{A}
\newcommand{\appa}{B}
\newcommand{\appc}{C}

\begin{document}

\title[Scattering matrix ensemble for time-dependent transport]{Scattering 
matrix ensemble for time-dependent transport through a chaotic quantum dot}

\author{M L Polianski and P W Brouwer}
\address{Laboratory of Atomic and Solid State Physics, Cornell
University, Ithaca, NY 14853-2501}

\date{\today}

\begin{abstract}
Random matrix theory can be used to describe the transport
properties of a chaotic quantum dot coupled to leads. In such a
description, two approaches have been taken in the literature,
considering either the Hamiltonian of the dot or its scattering
matrix as the fundamental random quantity of the theory. In this
paper, we calculate the first four moments of the distribution of
the scattering matrix of a chaotic quantum dot with a time-dependent
potential, thus establishing the foundations of a ``random
scattering matrix approach'' for time-dependent scattering. We
consider the limit that the number of channels $N$ coupling the
quantum dot the reservoirs is large. In that limit, the scattering
matrix distribution is almost Gaussian, with small non-Gaussian
corrections. Our results reproduce and unify results for
conductance and pumped current previously obtained in the
Hamiltonian approach. We also discuss an application to current
noise.

\end{abstract}

%

\pacs{73.23.-b, 72.10.Bg, 72.70.+m, 05.45.Mt\\
~\\
Submitted to: {\em J.\ Phys.\ A: Math.\ Gen.}}

\maketitle

\section{Introduction}

{}From a statistical point of view, energy levels and wavefunctions
in semiconductor quantum dots and metal grains, or
eigenfrequencies and eigenmodes of microwave cavities, share a
remarkable universality. With proper normalization, correlation
functions of energy levels or wavefunctions
for an ensemble of macroscopically equivalent, but
microscopically distinct samples depend on the fundamental
symmetries of the sample only; they do not depend on sample shape
or volume, or on the impurity concentration. 
The same universality appears for correlators of eigenvalues
and eigenfunctions of large matrices with randomly chosen
elements \cite{Mehta,MelloReview,GMW}.
Originally, such ``random matrices'' were introduced by Wigner
and Dyson to describe the universal features of spectral correlations
in heavy nuclei \cite{Porter,Pendry}.
Theoretical predictions from random matrix theory have
been verified in experiments on semiconductor quantum dots and
chaotic microwave cavities, and with the help of numerical
simulations \cite{Bohigas1,Bohigas2,Kouwenhoven,Alhassid}; 
for the case of a disordered quantum dot, the validity
of random matrix theory has been proven by field-theoretic
methods \cite{Efetovbook}.

Open samples, such as semiconductor quantum dots coupled to
source and drain reservoirs by means of ballistic
point contacts or microwave cavities coupled to ideal waveguides,
do not have well-resolved energy levels or wavefunctions.
They are characterized by means of a continuous density of states
and by their transport properties, such as
conductance or shot noise power. Within random-matrix theory,
two approaches have been taken to describe open samples
\cite{BeenakkerReview}. 
In both approaches, transport properties are described in
terms of the sample's scattering matrix ${\cal S}$.
The first approach is the ``random Hamiltonian approach''.
In this approach, the
scattering matrix is expressed in terms of a random Hermitean
matrix $H$, which represents the Hamiltonian of the closed sample.
Averages or fluctuations of transport properties are then
calculated in terms of the known statistical distribution of the
random matrix $H$ \cite{VWZ,LW}.
In the second approach, the ``random scattering matrix approach'',
the scattering matrix ${\cal S}$ itself is considered the fundamental random
quantity. It is taken from Dyson's ``circular ensemble'' of uniformly
distributed random unitary matrices 
\cite{BlumelSmilansky1,BlumelSmilansky2,BlumelSmilansky3,BM,JPB},
or a generalization known as the ``Poisson kernel'' 
\cite{MelloReview,MPS,Brouwer1995}.
Both approaches were shown to be equivalent
\cite{LW,Brouwer1995}. Hence, in the end, which method to use is a
matter of taste.

Recently, there has been interest in transport through chaotic
quantum dots
with a time-dependent Hamiltonian. Switkes {\em et al.} fabricated
a ``quantum electron pump'' consisting of a chaotic quantum
dot of which the shape could be changed by two independent
parameters \cite{Switkes}.
Periodic variation of the shape then causes
current flow through the quantum dot, hence the name ``electron
pump''. Motivated by theoretical predictions of Vavilov and Aleiner
\cite{VavilovAleiner,VA2},
Huibers {\em et al.} looked at the effect of microwave radiation on
the quantum interference corrections to the conductance of a
quantum dot \cite{Huibers}. The presence of a time-dependent
potential will cause the ratio of universal conductance
fluctuations with and without time-reversal symmetry
to be less than two if the typical frequency of the fluctuations
is of the order of the electron escape rate from the quantum
dot \cite{VA2,transport,transport1b,transport2}. 
(By the Dyson-Mehta theorem \cite{DysonMehta},
the ratio is two in the absence of a time-dependent perturbation
\cite{BeenakkerReview}.)

A scattering matrix formalism to describe time-dependent transport
was developed by B\"uttiker and coworkers
\cite{Buettiker1,Buettiker2,Buettiker3,BC,BJLTP}. 
The scattering matrix formalism
for time-dependent scattering is more complicated than the formalism
for time-independent scattering, since energy is no longer conserved
upon scattering from a cavity or quantum dot with a
time-dependent potential. In the adiabatic
limit, when the frequency $\omega$ of the
time-dependent variations is small compared to the escape rate
from the quantum dot, the theory can be formulated in terms
of the scattering matrix ${\cal S}$ and its derivative to energy
\cite{BC}. However, a theory that describes arbitrary frequencies
$\omega$ must be formulated
in terms of a scattering matrix ${\cal S}(\varepsilon,\varepsilon')$ that
depends on two energy arguments, or,
equivalently, a matrix ${\cal S}(t,t')$ depending
on two time arguments \cite{VavilovAleiner}.

Random matrix theory can be used to describe the statistics of
time-dependent transport if the time dependence is slow on the 
scale of the time $\tau_{\rm erg}$ needed for ergodic exploration
of the quantum dot.
In several recent papers 
\cite{VavilovAleiner,VA2,transport,transport1b,transport2,SAA,VAA}, 
the calculation of time-dependent transport
properties for an ensemble of chaotic quantum dots was done using a
variation of the
``random Hamiltonian approach'': the time-dependent scattering matrix 
${\cal S}(t,t')$ is first expressed in terms of a time-dependent Hermitean
matrix $H(t)$, which is the sum of a time-independent random matrix
and a time-dependent matrix which does not need to be random; 
the ensemble average is then calculated by
integrating $H$ over the appropriate distribution of random matrices.
It is the purpose of this paper to develop a ``random scattering 
matrix approach''
for time-dependent scattering, using the distribution of the 
scattering matrix
${\cal S}(t,t')$, not the Hamiltonian $H(t)$, as the starting point for
further calculations.

For time-independent scattering, the distribution of the elements
of the scattering matrix is given by the circular ensembles from
random matrix theory (or, for a quantum dot with non-ideal leads,
by the Poisson kernel). In the limit that the dimension $N$ of the
scattering matrix becomes large, the scattering matrix distribution
can be well approximated by a Gaussian, whereas
non-Gaussian correlations can be accounted 
for in a systematic expansion in $1/N$
\cite{unitary}. For the calculation of transport properties
(conductance, shot noise power), the Gaussian approximation
is usually sufficient; knowledge of the underlying ``full''
scattering matrix distribution is not required. Here, we
take a similar approach for time-dependent transport. We show that, 
for large $N$, elements
of the scattering matrix ${\cal S}(t,t')$ are almost Gaussian
random numbers, for which non-Gaussian correlations can be
taken into account by means of a systematic expansion in $1/N$.
We calculate the second moment of the distribution and the
leading non-Gaussian correction.

This paper is organized as follows:
In section \ref{sec:2a} we review the scattering matrix approach
for time-independent scattering. The case of time-dependent
scattering is considered in section \ref{sec:2}.
Applications are discussed in section \ref{sec:4}. Details
of the calculation and an extension to the case of quantum
dots with nonideal contacts can be found in the appendices.
The second moment of the scattering matrix distribution calculated 
here was used in reference \onlinecite{PVB}
to compute the shot noise power of a quantum electron pump.

\section{Time-independent scattering}
\label{sec:2a}

We first summarize important facts about the distribution of the
scattering matrix ${\cal S}$ for time-independent scattering.

For large matrix size $N$, the scattering matrix elements ${\cal S}_{ij}$
have a Gaussian distribution with small non-Gaussian correlations.
Mathematically, this is a consequence of the fact that ${\cal S}$ is
distributed according to the circular ensemble from random
matrix theory or, for a quantum dot with non-ideal
leads, the Poisson kernel \cite{unitary}. 
In a semiclassical picture, the
Gaussian distribution of the scattering matrix elements follows
from the central limit theorem, when ${\cal S}_{ij}$ is written as
a sum over many paths, where the contribution of each path
contains a random phase factor 
\cite{BlumelSmilansky1,BlumelSmilansky2,BlumelSmilansky3,JBS,Stone}. 
The small non-Gaussian corrections follow because the full scattering
matrix satisfies the constraint of unitarity, which is not
imposed in the semiclassical formulation.\footnote{Averages
or correlation functions of certain transport properties  
which, at first sight, would require knowledge of
the fourth moment of the scattering matrix distribution, can be formulated
in terms of the second moment only, using unitarity of the
scattering matrix. This way, the average and fluctuations of
the conductance of a chaotic quantum dot have been calculated
using the semiclassical approach, see, e.g., references 
\cite{Argaman,Vallejos}.}

The Gaussian part of the distribution is characterized by the
first two moments. In the main text, we focus on the case of
a quantum dot coupled to the outside world via
ideal leads. In this case, the first moment vanishes,
\begin{equation}
  \langle {\cal S}_{ij} \rangle = 0.
\end{equation}
The case of nonideal leads, for which $\langle {\cal S}_{ij} \rangle
\neq 0$, is discussed in appendix \appb.
The second moment of the scattering matrix distribution
depends on the presence or absence of time-reversal
symmetry (TRS),
\begin{eqnarray}
  W_1^{ij;kl} = \langle {\cal S}_{ij} {\cal S}_{kl}^* \rangle
  &=& {1 \over N} \times \left\{ \begin{array}{ll}
  ( \delta_{ik} \delta_{jl} + \delta_{il} \delta_{jk})
  & \mbox{with TRS}, \\
  \delta_{ik} \delta_{jl} &
  \mbox{without TRS},
  \end{array} \right.
  \label{eq:Sdistr}
\end{eqnarray}
up to corrections of relative order $1/N$ in the presence of
time-reversal symmetry.
All averages involving unequal powers of ${\cal S}$ and ${\cal S}^*$ 
vanish.
Equation (\ref{eq:Sdistr}) is for spinless particles or for electrons
with spin in the absence of spin-orbit coupling. (In the latter case, 
the scattering matrix has dimension $2N$ and is of the form ${\cal S} 
\otimes \openone_2$, where $\openone_2$ is the $2 \times 2$ unit 
matrix in spin space and the $N \times N$ matrix ${\cal S}$ describes
scattering between orbital scattering channels.) 
We do not consider the case of broken spin-rotation symmetry, when
${\cal S}$ is a random matrix of quaternions \cite{Mehta}.

Non-Gaussian
correlations of scattering matrices are of relative order $1/N$
or less. The leading non-Gaussian correlations are described
by the cumulant \cite{Samuel,Mello1990,unitary}
\begin{eqnarray}
  \fl
  W_2^{i_1j_1,i_2j_2;k_1l_1,k_2l_2} = \langle
  {\cal S}_{i_1j_1}
  {\cal S}_{i_2j_2}
  {\cal S}_{k_1l_1}^*
  {\cal S}_{k_2l_2}^*
  \rangle
  -
  \langle {\cal S}_{i_1j_1} {\cal S}_{k_1l_1}^* \rangle
  \langle {\cal S}_{i_2j_2} {\cal S}_{k_2l_2}^*
  \rangle 
  \nonumber \\ \mbox{} 
  -
  \langle {\cal S}_{i_1j_1} {\cal S}_{k_2l_2}^* \rangle
  \langle {\cal S}_{i_2j_2} {\cal S}_{k_1l_1}^*
  \rangle.
  \label{eq:W2indep}
\end{eqnarray}
In the absence of time-reversal symmetry and for $N \gg 1$, the
cumulant $W_2$ is given by
\begin{eqnarray}
  W_2 &=& - {1 \over N^3}
  \left(
  \delta_{i_1k_1} \delta_{j_1 l_2} \delta_{i_2 k_2} \delta_{j_2 l_1}
  +
  \delta_{i_1k_2} \delta_{j_1 l_1} \delta_{i_2 k_1} \delta_{j_2 l_2}
  \right),
  \label{eq:W2}
\end{eqnarray}
In the presence of time-reversal symmetry, $W_2$ is found by
the addition of 14 more terms to equation (\ref{eq:W2}),
corresponding to the permutations $i_2 \leftrightarrow j_2$,
$k_1 \leftrightarrow l_1$, and $k_2 \leftrightarrow l_2$,
\begin{eqnarray}
  \fl
  W_2 = - {1 \over N^3}
  \left( 
  \delta_{i_1 k_1} \delta_{j_1 l_2} \delta_{i_2 k_2} \delta_{j_2 l_1}
  +
  \delta_{i_1 k_2} \delta_{j_1 l_1} \delta_{i_2 k_1} \delta_{j_2 l_2}
  +
  \delta_{i_1 k_1} \delta_{j_1 l_2} \delta_{j_2 k_2} \delta_{i_2 l_1}
  +
  \delta_{i_1 k_2} \delta_{j_1 l_1} \delta_{j_2 k_1} \delta_{i_2 l_2}
  \right. \nonumber \\  \left. \mbox{}
  +
  \delta_{i_1 l_1} \delta_{j_1 l_2} \delta_{i_2 k_2} \delta_{j_2 k_1}
  +
  \delta_{i_1 k_2} \delta_{j_1 k_1} \delta_{i_2 l_1} \delta_{j_2 l_2}
  +
  \delta_{i_1 l_1} \delta_{j_1 l_2} \delta_{j_2 k_2} \delta_{i_2 k_1}
  \right. \nonumber \\  \left. \mbox{}
  +
  \delta_{i_1 k_2} \delta_{j_1 k_1} \delta_{j_2 l_1} \delta_{i_2 l_2}
  +
  \delta_{i_1 k_1} \delta_{j_1 k_2} \delta_{i_2 l_2} \delta_{j_2 l_1}
  +
  \delta_{i_1 l_2} \delta_{j_1 l_1} \delta_{i_2 k_1} \delta_{j_2 k_2}
  \right. \nonumber \\  \left. \mbox{}
  +
  \delta_{i_1 k_1} \delta_{j_1 k_2} \delta_{j_2 l_2} \delta_{i_2 l_1}
  +
  \delta_{i_1 l_2} \delta_{j_1 l_1} \delta_{j_2 k_1} \delta_{i_2 k_2}
  +
  \delta_{i_1 l_1} \delta_{j_1 k_2} \delta_{i_2 l_2} \delta_{j_2 k_1}
  \right. \nonumber \\  \left. \mbox{}
  +
  \delta_{i_1 l_2} \delta_{j_1 k_1} \delta_{i_2 l_1} \delta_{j_2 k_2}
  +
  \delta_{i_1 l_1} \delta_{j_1 k_2} \delta_{j_2 l_2} \delta_{i_2 k_1}
  +
  \delta_{i_1 l_2} \delta_{j_1 k_1} \delta_{j_2 l_1} \delta_{i_2 k_2}
  \right).
  \label{eq:W2TRS}
\end{eqnarray}
We refer to reference \onlinecite{unitary} for higher-order cumulants
and finite-$N$ corrections to $W_1$ and $W_2$.

Although equations (\ref{eq:Sdistr}) and (\ref{eq:W2indep}) do not
specify the full scattering matrix distribution --- for that one
would need to know all cumulants ---, they are sufficient to calculate
the average and variance of most transport properties.
As an example, we consider a quantum dot
connected to source and drain reservoirs by means of two
ballistic point contacts with $N_1$ and $N_2$ propagating
channels per spin direction at the Fermi level, with $N =
N_1 + N_2$. The
zero-temperature conductance is given by the Landauer
formula, which we write as \cite{Argaman}
\begin{equation}
  G = {2 e^2 \over h} \left( {N_1 N_2 \over N} - \tr
  {\cal S} \Lambda {\cal S}^{\dagger} \Lambda \right),
  \label{eq:Landauer}
\end{equation}
where $\cal S$ is the $N \times N$ scattering matrix and
$\Lambda$ is an $N \times N$ diagonal matrix with
\begin{equation}
  \Lambda_{ij} = {\delta_{ij} \over N} \times
  \left\{ \begin{array}{ll}
  N_2 & \mbox{if $1 \le i \le N_1$}, \\
  -N_1 & \mbox{if $N_1 < i \le N$}.
  \end{array} \right.
  \label{eq:Lambda}
\end{equation}
For large $N$, the second term in equation (\ref{eq:Landauer}) is a
small and fluctuating quantum correction to the classical
conductance of the quantum dot. Using equations (\ref{eq:Sdistr}) and
(\ref{eq:W2}), the average and variance of the conductance for $N
\gg 1$ then follow as
\begin{eqnarray}
  \langle G \rangle &=& {2 e^2 \over h}
  \left( {N_1 N_2 \over N} - \delta_{\beta,1}
  {N_1 N_2 \over N^2} \right) \ \
  \label{eq:Gavg} \\
  \mbox{var}\, G &=& {4 e^4 \over h^2}
  \left( {N_1^2 N_2^2 \over N^4} \right) (1 + \delta_{\beta,1}),
  \label{eq:varg}
\end{eqnarray}
where the symmetry parameter $\beta=1$ or $2$ with
or without time-reversal symmetry, respectively.

In the derivation of equations (\ref{eq:Gavg}) and (\ref{eq:varg})
it is important that
the matrix $\Lambda$ is traceless. This ensures that
the non-Gaussian cumulant (\ref{eq:W2}) does not contribute
to $\mbox{var}\, G$, despite the fact that calculation of
$\mbox{var}\, G$ involves an average over a product of four
scattering matrices. Similarly, the ${\cal O}(N^{-2})$ corrections
to the second moment $W_1$ of equation (\ref{eq:Sdistr})
in the presence of time-reversal symmetry do not contribute
to the average conductance to order $N^0$.

So far we have only considered elements of the scattering matrix
at one value of the Fermi energy $\varepsilon$ (and of the magnetic field,
etc.). If one wants to calculate averages involving scattering
matrices at different energies, one needs to know
the joint distribution of the scattering matrix ${\cal S}(\varepsilon)$ 
at different values of $\varepsilon$.
To date, no full solution to this problem is known for $N > 1$.
However, for large $N$, the joint distribution of scattering matrix
elements ${\cal S}_{ij}$ at different values of the Fermi energy or other
parameters continues to be well approximated by a Gaussian,
while unitarity causes non-Gaussian corrections that are small as
$1/N$. As before,
the Gaussian part of the distribution is specified by its first
and second moment. The first moment is zero for a quantum dot with
ideal leads; the second moment reads
\footnote{For the second moment, an exact solution was
obtained using the supersymmetry approach, see reference 
\onlinecite{VWZ}.}
\begin{eqnarray}
  W_1^{ij;kl}(\varepsilon;\varepsilon')
  &=&
  \langle {\cal S}_{ij}(\varepsilon)
  {\cal S}_{kl}(\varepsilon')^* \rangle
  \nonumber \\
  &=& {1 \over N - \rmi (\varepsilon - \varepsilon')}
\times \left\{ \begin{array}{ll}
  ( \delta_{ik} \delta_{jl} + \delta_{il} \delta_{jk})
  & \mbox{with TRS}, \\
  \delta_{ik} \delta_{jl} &
  \mbox{without TRS}.
  \end{array} \right.
  \label{eq:Sdistrparam} \label{eq:cum1}
\end{eqnarray}
Here, and below, we measure energy in units of $\Delta/2 \pi$,
where $\Delta$ is the mean spacing between the spin-degenerate
energy levels in the quantum dot without the leads.
Equation (\ref{eq:cum1}) was originally derived using semiclassical
methods
\cite{BlumelSmilansky1,BlumelSmilansky2,BlumelSmilansky3,JBS,Stone} 
and in the Hamiltonian approach of random-matrix
theory \cite{VWZ,Efetov95,Frahm95}. A derivation using the
random scattering matrix approach is given in reference
\cite{BB} and in appendix \appa.
In the absence of time-reversal symmetry, the leading
non-Gaussian correlations are described by the cumulant
\begin{eqnarray} \fl
  W_2^{i_1j_1,i_2j_2;k_1l_1,k_2l_2}
  (\varepsilon_1,\varepsilon_2;\varepsilon_1',\varepsilon_2')
  =
  \langle S^{\vphantom{*}}_{i_1 j_1}
  (\varepsilon_1) S^{\vphantom{*}}_{i_2 j_2}
  (\varepsilon_2) S^{*}_{k_1 l_1} (\varepsilon'_1)
  S^{*}_{k_2 l_2} (\varepsilon'_2)\rangle
  \nonumber \\ \mbox{}
  - \langle S^{\vphantom{*}}_{i_1 j_1} (\varepsilon_1)
  S^{*}_{k_1 l_1} (\varepsilon'_1) \rangle \langle
   S^{\vphantom{*}}_{i_2 j_2} (\varepsilon_2)
  S^{*}_{k_2 l_2} (\varepsilon'_2)\rangle
  \nonumber \\ \mbox{}
  - \langle S^{\vphantom{*}}_{i_1 j_1} (\varepsilon_1)
   S^{*}_{k_2 l_2} (\varepsilon'_2) \rangle \langle
   S^{\vphantom{*}}_{i_2 j_2} (\varepsilon_2)
   S^{*}_{k_1 l_1} (\varepsilon'_1)\rangle
  \nonumber \\ \lo
  =
  -{\left(
  \delta_{i_1k_1} \delta_{j_1 l_2} \delta_{i_2 k_2} \delta_{j_2 l_1}
  +
\delta_{i_1k_2} \delta_{j_1 l_1} \delta_{i_2 k_1} \delta_{j_2 l_2}
  \right)
  (N - \rmi (\varepsilon_1 + \varepsilon_2 -
    \varepsilon'_1 - \varepsilon'_2)) \over
  (N - \rmi (\varepsilon_1 - \varepsilon'_1))
  (N - \rmi (\varepsilon_1 - \varepsilon'_2))
  (N - \rmi (\varepsilon_2 - \varepsilon'_1))
  (N - \rmi (\varepsilon_2 - \varepsilon'_2))}.
  \label{eq:cum2}
\end{eqnarray}
In the presence of time-reversal symmetry, 14 terms corresponding
to the permutations
$i_2 \leftrightarrow j_2$, $k_1 \leftrightarrow l_1$, and
$k_2 \leftrightarrow l_2$ have to be added to equation
(\ref{eq:cum2}), respectively, as in equation (\ref{eq:W2TRS})
for the energy-independent case. A derivation
of equation (\ref{eq:cum2}) is given in appendix \appa.

Equations (\ref{eq:cum1}) and (\ref{eq:cum2}) can be used
to calculate averages and correlation functions for
transport properties that involve scattering matrices at
different energies.
As an example, using equation (\ref{eq:Sdistrparam}) for the second
moment of the scattering matrix distribution, the conductance
autocorrelation function is found as \cite{Efetov95,Frahm95}
\begin{eqnarray}
  \langle G(\varepsilon_1) G(\varepsilon_2) \rangle
  - \langle G(\varepsilon_1) \rangle
  \langle G(\varepsilon_2) \rangle  &=&
  {4 e^2 N_1^2 N_2^2 \over h^2 N^2}
  {(1 + \delta_{\beta,1}) \over N^2 + (\varepsilon_1 - \varepsilon_2)^2}.
  \label{eq:gcov}
\end{eqnarray}

\section{Time-dependent scattering} \label{sec:2}

For time-dependent scattering, the energies of incoming and
scattered particles do not need to be equal. In order to
describe scattering from a time-dependent scatterer, we use
a scattering matrix ${\cal S}(t,t')$ with two time arguments.
(We prefer to use the formulation with two time arguments
instead of a formulation in which ${\cal S}$ has two energy
arguments, since the former allows us to describe an arbitrary
time-dependence of the perturbations.)
For a quantum dot coupled to leads with, in total, $N$
scattering channels, the two-time scattering
matrix ${\cal S}(t,t')$ relates the annihilation
operators ${\bf a}_{i}(t)$ and ${\bf b}_{i}(t)$ of
incoming states and outgoing states in channel $i=1,\ldots,N$,
\begin{eqnarray}
{\bf b}_{i}(t)&=& \sum_{j=1}^{N}
  \int_{-\infty}^{+\infty} {\cal S}_{ij}(t,t')
  {\bf a}_{j}(t')dt', \nonumber \\
  {\bf b}_{i}^{\dagger}(t)&=& \sum_{j=1}^{N} \int_{-\infty}^{+\infty}
  {\bf a}_{j}^\dagger(t')
  ({\cal S}^\dagger(t',t))_{ji}dt'.
\label{eq:Sdef}
\end{eqnarray}
Causality imposes that
\begin{eqnarray}
  && {\cal S}(t,t') = 0\ \ \mbox{if $t < t'$}.
\end{eqnarray}
Unitarity is ensured by the condition
\begin{eqnarray}
  && \sum_{j=1}^{N} 
    \int dt ({\cal S}^{\dagger}(t'',t))_{ij} {\cal S}_{jk}(t,t')
  = \delta(t''-t') \delta_{ik}, \nonumber \\
  && \sum_{j=1}^{N} \int dt {\cal S}_{ij}(t'',t) 
    ({\cal S}^{\dagger}(t,t'))_{jk}
  = \delta(t''-t') \delta_{ik},
  \label{eq:unitarity}
\end{eqnarray}
where the Hermitean conjugate scattering matrix
${\cal S}^{\dagger}(t',t)$ is defined as
\begin{eqnarray}
\left({\cal S}^\dagger(t',t)\right)_{ij}=
{\cal S}_{ji}^*(t,t').
\end{eqnarray}

For a quantum dot without time-independent potential, the scattering 
matrix ${\cal S}^{0}(t,t')$
depends on the difference $t-t'$ only. 
(In this section, we use a superscript ``$0$'' to indicate that 
${\cal S}^{0}$ is a scattering matrix for time-independent scattering.)
It is related the scattering
matrix in energy representation by Fourier transform,
\begin{equation}
  {\cal S}^{0}(t,t') = {1 \over \hplanck}
  \int_{-\infty}^\infty d\varepsilon \,
  {\cal S}^{0}(\varepsilon)\rme^{i \varepsilon(t-t')/\hbar}.
\end{equation}
Borrowing results from the previous section, we infer that the
elements of ${\cal S}^{0}(t,t')$ have a distribution that is almost
Gaussian --- the Fourier transform of a Gaussian is a Gaussian 
as well ---, but with non-Gaussian correlations that are small as
$N \to \infty$. Fourier transforming equation (\ref{eq:cum1}), we
obtain the variance of the distribution \cite{ABG}
\begin{eqnarray}  \fl
  \langle {\cal S}^{0\vphantom{*}}_{i j}
  (t,t')
  {\cal S}^{0*}_{k l} (s,s')\rangle
  = \delta(t-t'-s+s') \theta(t-t')
  {\cal D}^0(t-t')
  \nonumber \\ \mbox{} 
  \times \left\{ \begin{array}{ll}
  ( \delta_{ik} \delta_{jl} + \delta_{il} \delta_{jk})
  & \mbox{with TRS}, \\
  \delta_{ik} \delta_{jl} &
  \mbox{without TRS}.
  \end{array} \right.
  \label{eq:cum1t}
\end{eqnarray}
Here, time is measured in units of $\hplanck/\Delta$
and the function ${\cal D}^0$ is given by
\begin{equation}
  {\cal D}^0(\tau) = \rme^{-N \tau}.
\end{equation}
Fourier transform of equation (\ref{eq:cum2}) gives the leading
non-Gaussian contribution,
\begin{eqnarray} \fl
  W_2^{0;i_1 j_1,i_2 j_2;k_1 l_2,k_2 l_2}
  (t_1,t_1';t_2,t_2';s_1,s_1';s_2,s_2')
   \nonumber \\ \lo
  =
  \langle {\cal S}^{0\vphantom{*}}_{i_1 j_1}
  (t_1,t_1') {\cal S}^{0\vphantom{*}}_{i_2 j_2}
  (t_2,t_2') {\cal S}^{0*}_{k_1 l_1} (s_1,s_1')
  {\cal S}^{0*}_{k_2 l_2} (s_2,s_2')\rangle
  \nonumber \\ \mbox{}
  -
  \langle {\cal S}^{0\vphantom{*}}_{i_1 j_1}(t_1,t_1')
  {\cal S}^{0*}_{k_1 l_1} (s_1,s_1') \rangle
  \langle
  {\cal S}^{0\vphantom{*}}_{i_2 j_2}
  (t_2,t_2')
  {\cal S}^{0*}_{k_2 l_2} (s_2,s_2')\rangle
  \nonumber \\ \mbox{}
  -
  \langle {\cal S}^{0\vphantom{*}}_{i_1 j_1}
  (t_1,t_1') {\cal S}^{0*}_{k_2 l_2} (s_2,s_2')
  \rangle \langle
  {\cal S}^{0\vphantom{*}}_{i_2 j_2}
  (t_2,t_2') {\cal S}^{0*}_{k_1 l_1} (s_1,s_1')
  \rangle
  \label{eq:cum2t}
  \\ \lo =
  \left(
  \delta_{i_1k_1} \delta_{j_1 l_2} \delta_{i_2 k_2} \delta_{j_2 l_1}
  +
  \delta_{i_1k_2} \delta_{j_1 l_1} \delta_{i_2 k_1} \delta_{j_2 l_2}
  \right)
  {\cal F}^{0}(t_1-t_1';t_2-t_2';s_1-s_1';s_2-s_2')
  \nonumber \\ \mbox{} \times
  \delta(t_1-t_1'+t_2-t_2'-s_1+s_1'-s_2+s_2') 
  \theta(t_1-t_1') \theta(t_2-t_2') \theta(s_1-s_1'),
  \nonumber
\end{eqnarray}
with
\begin{eqnarray} \fl
  {\cal F}^0(\tau_1;\tau_2;\tau_3;\tau_4) &=&
  [N \min(\tau_1,\tau_2,\tau_3,\tau_4)-1]
  \rme^{-N(\tau_1+\tau_2)} .
\end{eqnarray}
Note that, in view of the delta function in equation (\ref{eq:cum2t}),
the function ${\cal F}^{0}$ depends on three time variables only.
Despite the redundancy, we keep the four time arguments for
notational convenience.
As before, in the presence of time-reversal symmetry, the
expression for the cumulant is obtained
by adding terms that are obtained after interchanging
$i_2 \leftrightarrow j_2$, $k_1 \leftrightarrow l_1$, and
$k_2 \leftrightarrow l_2$, cf.\
Eq.\ (\ref{eq:W2TRS}).

In order to calculate the defining cumulants $W_1$ and $W_2$ for
the case of a chaotic quantum dot with a time-dependent potential,
we need a statistical model for the scattering matrix distribution
for time-dependent scattering. Such a model can be provided by
the Hamiltonian approach \cite{VavilovAleiner}, or, alternatively,
by extending the ``stub model'' of references
\onlinecite{BB,BrouwerBeenakker96,waves} to the case of
time-dependent scattering{\footnote{The stub model is similar 
in spirit to the ``quantum graph'', the spectral
        statistics of which is known to follow
        random-matrix theory \cite{Kottos}.}.
In the latter approach,
the $N \times N$ scattering matrix
${\cal S}(t,t')$ is written in terms of an $M \times M$ random
matrix ${\cal U}(t,t')$ (with $M \gg N$) and a $(M-N) \times (M-N)$
random Hermitean matrix $H$,
\begin{eqnarray}\label{eq:stub}
{\cal S}=P{\cal U}(1-R {\cal U})^{-1}P^\dagger,\ \
  R = Q^{\dagger} \rme^{-2 \pi \rmi H/M \Delta} Q.
  \label{eq:RH}
\end{eqnarray}
Here $P$ is an $N \times M$ matrix with $P_{ij}=\delta_{i,j}$
and $Q$ is an $(M-N)\times M$ matrix with $Q_{ij}=
\delta_{i+N,j}$. The scattering matrices ${\cal S}(t,t')$ and
${\cal U}(t,t')$ depend on two time indices, and the matrix
products involving ${\cal U}(t,t')$ in equation (\ref{eq:stub}) also
imply integration over intermediate times. The Hermitean matrix
$H$ depends on a single time argument and models both time-independent
and time-dependent perturbations to the Hamiltonian of the quantum
dot. 
The matrix ${\cal U}(t,t')$
depends on the time difference $t-t'$ only and satisfies the
constraint of unitarity, equation (\ref{eq:unitarity}) above. 
As the effect of a time-reversal symmetry breaking
magnetic field will be included
in $H$, cf.\ equation (\ref{eq:Hmagn}) below,
we further require that the matrix ${\cal U}$ 
is time-reversal symmetric,
\begin{equation}
  {\cal U}_{ij}(t-t') =
  {\cal U}_{ji}(t-t').
\end{equation}
The statistical distribution of the matrix ${\cal U}$ 
is the same as that of the scattering matrix of a
chaotic quantum dot coupled to a lead with $M$ channels, but without
magnetic field and time-dependent potential. Hence, the first
nonvanishing moments of the distribution are given by equations
(\ref{eq:cum1t}) and (\ref{eq:cum2t}) above, with ${\cal S}^{0}$
replaced by ${\cal U}$ and $N$ by $M$.

The physical idea behind equation (\ref{eq:stub}) is that the
time-dependent part of the potential is located in a ``stub'' (a
closed lead), see figure \ref{fig:stub}. The number of channels in
the stub is $M-N$. The matrix ${\cal U}$ is the $M \times M$
scattering matrix of the quantum dot without the stub; the scattering
matrix ${\cal S}$ is the scattering matrix of the entire system
consisting of the dot and the stub, 
taking into account the time-dependent scattering
from the stub. The matrix $R$ represents the time-dependent
scattering matrix for scattering from the stub. The stub is chosen
to be small compared to the quantum dot, so that
reflection from the stub can be regarded instantaneous --- that's
why the matrix $R(t)$ depends on a single time argument only. At
the end of the calculation, we take the limit $M \to \infty$. This
limit
ensures that the dwell time in the dot, which is proportional
to $1/N$, is much larger than the time of ergodic
exploration of the dot-stub system, which is proportional to
$1/M$. It is only in this limit that the scattering matrix
acquires a universal distribution which is described by random
matrix theory. Once the limit $M \to \infty$ is taken, the spatial
separation of chaotic scattering (described by the $M \times M$
scattering matrix ${\cal U}$) and the interaction with the
time-dependent potential (described by the time-dependent
reflection matrix $R$) no longer affects the distribution of the
scattering matrix ${\cal S}$ and the scattering matrix
distribution found using the stub model becomes identical to that
with a spatially distributed time-dependent potential in the
Hamiltonian approach.

A similar model has been used to describe the parametric
dependence of the scattering matrix in the scattering matrix
approach \cite{BB,BrouwerBeenakker96,BCH}. For the
parametric dependence of ${\cal S}$, equivalence
of the ``stub'' model and the Hamiltonian approach was shown
in reference \onlinecite{waves}. The calculational advantage
of the ``stub'' model is that, for a quantum dot with ideal
leads, the vanishing of the first moment $\langle {\cal S}_{ij}
\rangle = 0$ is manifest throughout the calculation, while it
requires fine-tuning of parameters at the end of the 
calculation in the Hamiltonian approach.

\begin{figure}
\epsfxsize=0.55\hsize
\hspace{0.25\hsize}
\epsffile{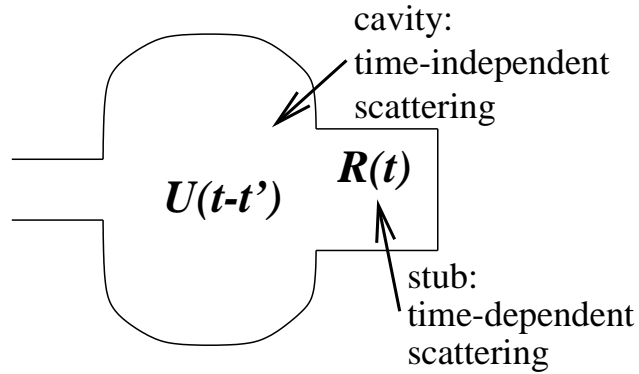}
\caption{\label{fig:stub} Cartoon of the picture behind
equation (\protect\ref{eq:stub}): Scattering from the chaotic quantum
dot with
the time-dependent potential is modeled as scattering from a
chaotic quantum dot with time-independent potential and a stub with
time-dependent potential.}
\end{figure}

The matrix $H$ in equation (\ref{eq:RH}) can be written as a sum
of three terms, describing three different perturbations to the
Hamiltonian of the quantum dot\footnote{
In the Hamiltonian approach, the parameters $x_1(t)$, $x_2(t)$,
$V(t)$, and $\alpha(t)$ of equations (\ref{eq:Hsum})--(\ref{eq:alpha})
correspond to time-dependent variations of the form
$$
  {\cal H}(t) =
  S + V(t) \openone +
  \frac{\rmi}{\sqrt{2M}} \alpha(t) A +
  \sum_{j=1}^{n} \frac{1}{\sqrt{M}} x_j(t) X_j,
$$
where $S$ and $X_j$ are real symmetric random $M \times M$
matrices, $j=1,\ldots,n$, $A$ is a
real antisymmetric random $M \times M$
matrix, and $\openone$ is the $M \times M$ unit matrix.
The off-diagonal elements of these
random matrices are Gaussian random numbers with zero mean and
unit variance. The diagonal elements of $S$ and $X_j$ have twice
the variance of the off-diagonal elements.}
\begin{equation}
  H = V(t) \openone + H_{\rm shape} + H_{\rm magn}. \label{eq:Hsum}
\end{equation}
The first term in equation (\ref{eq:Hsum}) represents an overall
shift of the potential $V(t)$ in the quantum dot. The second term
represents the effect of a variation of the shape of the
quantum dot,
\begin{equation}\label{eq:Hamiltonian}
  H_{\rm shape}(t) = \sum_{j=1}^{n} x_j(t) {X_j \Delta \over \pi}
\end{equation}
Here the $x_j$ ($j=1,\ldots,n$) are $n$ time-dependent
parameters governing the shape of
the quantum dot, and the $X_j$ are real symmetric random
$(M-N)\times (M-N)$ matrices with $\tr X_i X_j = M^2 \delta_{ij}$,
$i,j=1,\ldots,n$. Having more than one parameter to characterize
the dot's shape is important for applications to quantum pumping
\cite{Switkes,spivak,B_p,ZSA}.
The third term in equation (\ref{eq:Hsum}) represents the
parametric dependence
of the Hamiltonian on a magnetic flux $\Phi$ through the quantum dot
\begin{equation}
  H_{\rm magn}(t) = \rmi \alpha(t) {A \Delta \over \pi \sqrt{2}},
  \label{eq:Hmagn}
\end{equation}
where $A$ is a random antisymmetric $(M-N)\times (M-N)$ matrix
with $\tr A^{\rm T} A = M^2$. For a dot with diffusive electron
motion (elastic mean free path $l$, dot size $L \gg l$) one has
\begin{equation}
  \alpha^2 = \kappa \left( \frac{e \Phi(t)}{hc} \right)^2
  \frac{\hbar v_F l}{L^2 \Delta},
  \label{eq:alpha}
\end{equation}
where $\kappa$ is a constant of order unity and $\Phi$ the flux
through the quantum dot. One has $\kappa=4\pi/15$
for a diffusive sphere of radius $L$ and $\kappa=\pi/2$ for
a diffusive disk of radius $L$ \cite{FrahmPichard}.
For ballistic electron motion
with diffusive boundary scattering, the mean free path $l$
in equation (\ref{eq:alpha}) is replaced by $5L/8$ and
$\pi L/4$ for the cases of a sphere and a disk, 
respectively. (For the ballistic case, the value of 
$\alpha^2$ reported in reference \onlinecite{FrahmPichard}
is incorrect, see reference \onlinecite{APWB}.) In order to
ensure the validity of the random matrix theory, the time
dependence of the parameters $x_j$ and $\alpha$ should be
slow on the scale of the ergodic time $\tau_{\rm erg}$ of
the quantum dot.

Note that the description (\ref{eq:stub})--(\ref{eq:alpha})
contains the dependence on
a magnetic field explicitly. Having the full dependence on the
magnetic field at our disposal, we no longer need to distinguish
between the cases of presence and absence of time-reversal
symmetry.

Expanding equation (\ref{eq:RH}) in powers of $R$, the scattering
matrix ${\cal S}$ is calculated as a sum over ``trajectories'' that
involve chaotic scattering in the quantum dot and reflections
from the stub. Since 
different ``trajectories'' 
involve different channels
in the stub at different times, each term 
in the expansion carries a random phase, determined
by the random phases of the elements of ${\cal U}$. Hence,  elements
${\cal S}_{ij}$ will have a distribution that is almost Gaussian
for large $N$, since they are sums
over many contributions with random phases. 
Unitarity, imposed by the constraint
(\ref{eq:unitarity}) for the matrix ${\cal U}$ and the form
of the matrices ${\cal S}$ and $R$ in equation (\ref{eq:stub}) $R$,
leads to corrections to the Gaussian
distribution that are small as $N$ becomes large. 

The Gaussian
part of the distribution of the time-dependent scattering
matrix ${\cal S}(t,t')$ is specified by the second moment,
\begin{equation}
  W_1^{ij;kl}(t,t';s,s') =
  \langle {\cal S}^{\vphantom{*}}_{ij}
  (t,t')
  {\cal S}^{*}_{kl} (s,s')\rangle,
  \label{eq:W1t}
\end{equation}
whereas the leading non-Gaussian corrections are described by the
cumulant
\begin{eqnarray}
\fl
  W_2^{i_1j_1,i_2j_2;k_1l_1,k_2l_2}(t_1,t_1';
  t_2,t_2';s_1,s_1';s_2,s_2') \nonumber \\ \lo =
  \langle
  {\cal S}_{i_1j_1}(t_1,t_1')
  {\cal S}_{i_2j_2}(t_2,t_2')
  {\cal S}_{k_1l_1}^*(s_1,s_1')
  {\cal S}_{k_2l_2}^*(s_2,s_2')
  \rangle
  \nonumber \\ \mbox{} -
  \langle {\cal S}_{i_1j_1}(t_1,t_1') {\cal S}_{k_1l_1}^*(s_1,s_1') \rangle
  \langle {\cal S}_{i_2j_2}(t_2,t_2') {\cal S}_{k_2l_2}^*(s_2,s_2')
  \rangle \nonumber \\ \mbox{} -
  \langle {\cal S}_{i_1j_1}(t_1,t_1') {\cal S}_{k_2l_2}^*(s_2,s_2') \rangle
  \langle {\cal S}_{i_2j_2}(t_2,t_2') {\cal S}_{k_1l_1}^*(s_1,s_1')
  \rangle.
  \label{eq:W2tdef}
\end{eqnarray}

\begin{figure}
\epsfxsize=0.8\hsize
\hspace{0.15\hsize}
\epsffile{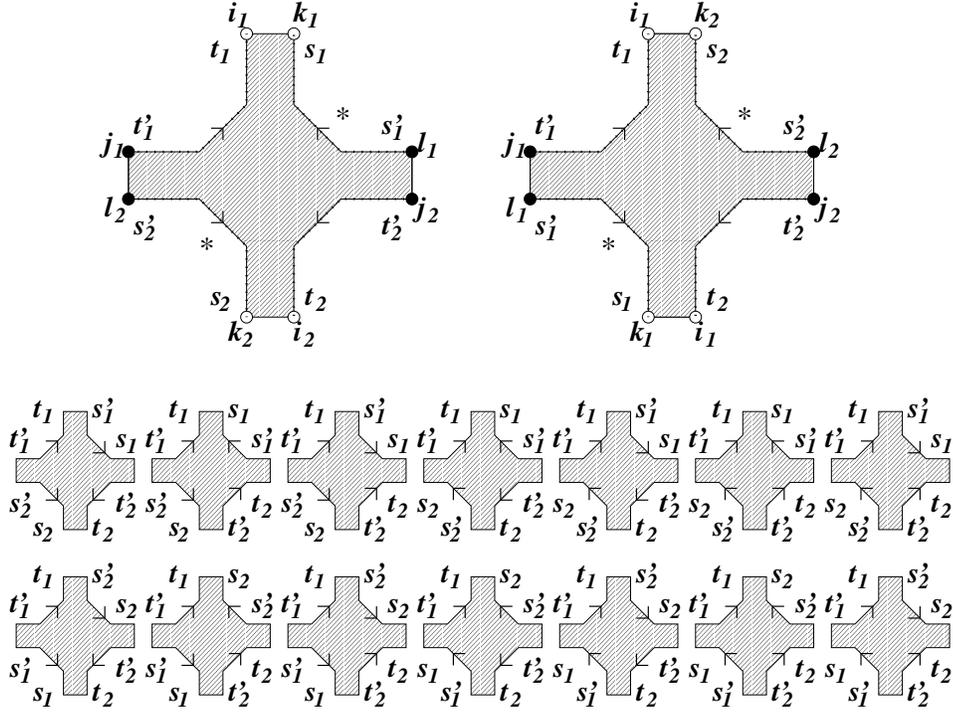}

\caption{\label{4corrdef}
Diagrammatic representation of the contributions to the
leading non-Gaussian correlator $W_2$ of equation (\protect\ref{eq:W2tdef}), 
which involves four scattering matrices. The top left
diagram has weight ${\cal F}(t_1,t_1';t_2,t_2';s_1,s_1';s_2,s_2')$,
cf.\ equation (\protect\ref{eq:W2t}). The top right diagram is 
obtained by interchange of two
scattering matrices and has weight
${\cal F}(t_1,t_1';t_2,t_2';s_2,s_2';s_1,s_1')$. These two diagrams
give all contributions to the cumulant $W_2$ in the absence
of time-reversal symmetry. In the presence of time-reversal
symmetry, the fourteen lower diagrams, corresponding to the reversal
of one or more directions of the vertices, contribute as well.}
\end{figure}

The central result of this paper is a calculation of the cumulants 
$W_1$ and $W_2$ for time-dependent scattering. 
Details of the calculation are reported in  appendix \appc.
For the second moment $W_1$ we find
\begin{eqnarray}
  \fl
  W_1^{ij;kl}(t,t';s,s') &=& \delta(t-t'-s+s')
  \theta(t-t')
  \left[ \delta_{ik} \delta_{jl}
  {\cal D}(t,t';s,s')
  + \delta_{il} \delta_{jk}
  {\cal D}(t,t';s',s) \right],
  \label{eq:W2D}
\end{eqnarray}
with
\begin{eqnarray}
  \fl
  \label{eq:Dresult}
  {\cal D}(\tau,\sigma;\tau', \sigma')
  =
  \exp\left[-N |\tau-\sigma| - {2 \pi \over \Delta}
  \int_{0}^{|\tau-\sigma|} d \xi \left(V(\sigma+\eta\xi) -
  V(\sigma'+\eta'\xi) \right)
  \right.
  \\ \nonumber \left. \mbox{} + 
  2 \sum_{j} \left[x_j(\sigma+\eta\xi) - x_j(\sigma'+\eta'\xi)\right]^2
  + \left[\eta \alpha(\sigma+\eta\xi) -  \eta'\alpha(\sigma'+\eta'\xi)
  \right]^2\right],
\end{eqnarray}
and $\eta =\sign(\tau-\sigma)$, $\eta' =\sign(\tau'-\sigma')$.
The first term in equation (\ref{eq:W2D})
is the analogue of the diffuson from standard
diagrammatic perturbation theory, while the second term 
corresponds to the cooperon. For notational convenience, both
terms are denoted by the same symbol ${\cal D}$. (Note that the 
order of the time arguments $s$ and $s'$ is reversed in the second
term of equation (\ref{eq:W2D}).)
The leading non-Gaussian
corrections are given by the cumulant $W_2$ for which we find
\begin{eqnarray}
  \fl
  W_2^{i_1j_1,i_2j_2;k_1l_1,k_2l_2}(t_1,t_1';
  t_2,t_2';s_1,s_1';s_2,s_2')
  =
  \theta(t_1-t_1') \theta(t_2-t_2') \theta(s_1-s_1')
  \theta(s_2-s_2')
  \nonumber \\ \mbox{} \times
  \delta_{i_1k_1} \delta_{j_1 l_2} \delta_{i_2 k_2} \delta_{j_2 l_1}
  \delta(t_1-t_1'+t_2-t_2'-s_1+s_1'-s_2+s_2')
  \nonumber \\ \mbox{} \times
  {\cal F}(t_1,t_1';t_2,t_2';s_1,s_1';s_2,s_2')
  + \mbox{permutations}.
  \label{eq:W2t}
\end{eqnarray}
The ``permutations'' in equation (\ref{eq:W2t}) refer to one term 
corresponding to the permutation $(s_1,s_1',k_1,l_1)
\leftrightarrow (s_2,s_2',k_2,l_2)$ of the third and fourth
arguments of $W_2$ and fourteen more
terms corresponding to the interchange of incoming and
outgoing channel and time arguments within the second, third, and
fourth argument of $W_2$. A diagrammatic representation of the
cumulant (\ref{eq:W2t}) and the relevant perturbations is shown in
figure \ref{4corrdef}.
The kernel ${\cal F}$ reads
\begin{eqnarray} \fl
  {\cal F}(\tau_1,\sigma_1;\tau_2,\sigma_2;
  \tau_1',\sigma_1';\tau_2',\sigma_2') =
  \int d\xi\,
  {\cal D}(\sigma_1+\eta_1\xi,\sigma_1;\sigma_1'+\eta_1'\xi,\sigma_1')
  \nonumber \\ \mbox{} \times
  {\cal D}(\tau_2-\tau_1'+\sigma_1'+\eta_2\xi,\sigma_2;
    \tau_2'-\tau_1+\sigma_1+\eta_2'\xi,\sigma_2')
  \nonumber \\  \mbox{} \times
  {\cal D}(\tau_1,\sigma_1+\eta_1\xi;
    \tau_2',\tau_2'-\tau_1+\sigma_1+\eta_2'\xi)
  \nonumber \\ \mbox{} \times
  {\cal D}(\tau_2,\tau_2-\tau_1'+\sigma_1'+\eta_2\xi;
    \tau_1',\sigma_1'+\eta_1'\xi)
  \nonumber \\ \mbox{} \times
  \left\{ N + 
  4\sum_{m}[x_m(\sigma_1+\eta_1\xi) - x_m(\tau_2'-\tau_1+\sigma_1+\eta_2'\xi)]
  \right. \nonumber \\ \left. \ \ \mbox{} \times
  [x_m(\sigma_1'+\eta_1'\xi)- x_m(\tau_2-\tau_1'+\sigma_1'+\eta_2\xi)]
  -\delta(\xi-|\tau_1-\sigma_1|) 
  \right. \nonumber \\ \ \ \left. \mbox{}
  - \delta(\xi-|\tau_1'-\sigma_1'|)
  + 2 [\eta_1\alpha(\sigma_1+\eta_1\xi) -
  \eta_2'\alpha(\tau_2'-\tau_1-\sigma_1+\eta_2'\xi)]
  \right. \nonumber \\ \left. \ \ \mbox{} \times
  [\eta_1'\alpha(\sigma_1'+\eta_1'\xi)-
  \eta_2\alpha(\tau_2-\tau_1'+\sigma_1'+\eta_2\xi)]
  \vphantom{\sum_m} \right\},
  \label{eq:F}
\end{eqnarray}
where we abbreviated $\eta_1 = \sign(\tau_1-\sigma_1)$,
$\eta_2 = \sign(\tau_2-\sigma_2)$, $\eta_1' = \sign(\tau_1'-\sigma_1')$,
and $\eta_2' = \sign(\tau_2'-\sigma_2')$. Note that equations
(\ref{eq:W2D})--(\ref{eq:F}) cover both the cases with and 
without time-reversal symmetry through 
the explicit dependence on the magnetic flux $\alpha$. 
If time-reversal symmetry 
is fully broken, all permutations in equation (\ref{eq:W2t}) that
involve the interchange of incoming and outgoing channels,
corresponding to the fourteen lower diagrams in figure
\ref{4corrdef}, vanish, and only the first two diagrams in
figure \ref{4corrdef} remain. Partial integration of the intermediate
time $\xi$ allows one to rewrite terms between brackets
$\{ \ldots \}$ in equation (\ref{eq:F}), see appendix
\appc\ for details. Finally, one verifies that the result
(\ref{eq:cum2t}) is recovered for $\alpha = 0$ and $\alpha \gg 1$, 
corresponding to presence and absence of time-reversal symmetry,
when the parameters $x_j$ and $\alpha$ do not depend on time.

\section{Applications} \label{sec:4}

In order to illustrate the use of equations (\ref{eq:W2D})--(\ref{eq:F}),
we return to the example of section \ref{sec:2a} and
consider transport through
a chaotic quantum dot coupled to two electrons
reservoirs by means of ballistic point contacts with $N_1$ and
$N_2$ channels, respectively. The scattering
matrix of the quantum dot has dimension $N=N_1+N_2$.
The current through the dot
is defined as a linear combination of the of the currents through
the two point contacts,
\begin{eqnarray}
\label{eq:current}
  {\bf I}(t) &=&
  e \sum_{i,j=1}^{N}
  \left({\bf a}_{i}^\dagger(t){\Lambda}_{ij}
  {\bf a}_{j}(t)-
  {\bf b}_{i}^\dagger(t)\Lambda_{ij}
  {\bf b}_{j}(t)\right),
\end{eqnarray}
where the $N \times N$ matrix $\Lambda$ was defined in
equation (\ref{eq:Lambda}) and the operators ${\bf a}_{i}(t)$
and ${\bf b}_{i}(t)$ are annihilation operators for incoming
and outgoing states in channel $i=1,\ldots,N$ in the leads, 
respectively, see section \ref{sec:2}.
The advantage of the definition (\ref{eq:current}) for the 
current through the quantum dot, instead of a definition
where the current through one of the contacts is used, is
that it simplifies the ensemble average taken below.
Both definitions of the current give the same result
for the quantity of interest, the integral of ${\bf I}(t)$ over
a large time interval $t_{\rm i} < t < t_{\rm f}$.

The electron distribution function for the electrons entering
the quantum dot from the leads is given
by the Fourier transform $f(t)$ of the Fermi function in the
corresponding electron reservoir \cite{scatt1,scatt2,scatt3},
\begin{eqnarray}
  \overline{{\bf a}_{j}^\dagger(t')
  {\bf a}_{i}^{\vphantom{\dagger}}(t)}
  &=& f_{ij}(t'-t),
  \nonumber \\
  \overline{{\bf a}_{j}^{\vphantom{\dagger}}(t')
  {\bf a}_{i}^{\dagger}(t)}
  &=& \tilde f_{i j}(t-t'),
  \label{eq:aaverage}
\end{eqnarray}
where we defined
\begin{eqnarray}
\label{eq:fdistribution}
f_{ij}(t) &=& \delta_{ij}
  \int {d \varepsilon \over 2 \pi \hbar}
 \frac{\rme^{\rmi\varepsilon t/\hbar}}
{\rme^{(\varepsilon-\mu_{i})/kT}+1}
  \nonumber \\ &=& \delta_{ij}
\frac{ikT \rme^{\rmi \mu_{i} t/\hbar}}{2 \hbar \sinh(\pi k T t/\hbar)},
  \nonumber \\
  \tilde f_{ij}(t)
  &=& \delta_{ij}\delta(t) - f_{ij}(t).
\end{eqnarray}
Here $\mu_{i}$ is the chemical potential of reservoir $1$ for
$1 \le i \le N_1$ and the chemical potential of reservoir $2$ for
$N_1 < i \le N$.

Substitution of equations (\ref{eq:aaverage}) and (\ref{eq:Sdef}) into
equation (\ref{eq:current}) allows us to calculate the time-averaged
expectation value of the current through the quantum dot for a
time interval $t_{\rm i} < t < t_{\rm f}$,
\begin{eqnarray}
  \fl
  I = {2e \over t_{\rm f} - t_{\rm i}}
  \int_{t_{\rm i}}^{t_{\rm f}} dt
  \int dt_1 dt_2
  \tr \left[ \delta(t-t_1)
  \Lambda \delta(t-t_2) -
  {\cal S}^{\dagger}(t_1,t) \Lambda
  {\cal S}(t,t_2) \right] f(t_1-t_2).
  \label{eq:currentgeneral}
\end{eqnarray}
(A factor two has been added to account for spin degeneracy.
The time interval $t_{\rm i} < t < t_{\rm f}$ during which charge
is measured is taken to be the largest time scale in the problem.)

In the absence of a source-drain voltage, equation
(\ref{eq:currentgeneral}) describes
the current that is ``pumped'' by the
time-dependent potential in the dot,
\begin{eqnarray}
  I_{\rm pump} &=&
  -{2e \over t_{\rm f} - t_{\rm i}}
  \int_{t_{\rm i}}^{t_{\rm f}} dt
  \int dt_1 dt_2\tr
  {\cal S}^{\dagger}(t_1,t) \Lambda
  {\cal S}(t,t_2) f_{\rm eq}(t_1-t_2),
  \label{eq:ipumpVAA}
\end{eqnarray}
where $f_{\rm eq}$ is the Fourier transform of the Fermi function.
Equation (\ref{eq:ipumpVAA}) was first derived in reference
\onlinecite{VAA}; it reduces to the current formulae of references
\onlinecite{B_p} and \onlinecite{ZSA} in the adiabatic
limit, where the time dependence of the potential of the quantum
dot is slow compared to the dwell time in the quantum dot.
At small bias voltage, there is a current proportional to the bias,
$I = G V$, where $G$ is the (time-averaged) conductance of the
dot. The conductance $G$ can be calculated from
equation (\ref{eq:currentgeneral})
by setting $\mu_{i} = \Lambda_{ii} e V$ and then
linearizing in $V$ \cite{VavilovAleiner},
\begin{eqnarray}
  \label{eq:conductance}
  \fl
  G &=& {2 e^2 \over h}
  \left[ {N_1 N_2 \over N}
  -
  {2 \pi \rmi \over t_{\rm f} - t_{\rm i}}
  \int_{t_{\rm i}}^{t_{\rm f}} dt
  \int dt_1 dt_2 (t_1-t_2) \tr \Lambda {\cal S}(t,t_1)
  \Lambda {\cal S}^{\dagger}(t_2,t)
  f_{\rm eq}(t_1-t_2) \right].
\end{eqnarray}
Here $f_{\rm eq}$ is the Fermi function in the absence of
the external bias.
For time-independent transport, equation (\ref{eq:conductance})
is equal to the Landauer formula (\ref{eq:Landauer}).

{\em Conductance.}
The ensemble average and the variance of the conductance
$G$ for a quantum dot
with a shape depending on a single time-dependent parameter
$x$ was calculated by
Vavilov and Aleiner using the Hamiltonian approach
\cite{VavilovAleiner}. Using the scattering matrix
correlator (\ref{eq:W1t}), their result for
$\langle G \rangle$ is easily reproduced and generalized
to arbitrary values of the (time-independent) magnetic field,
\begin{eqnarray}
  \fl \label{eq:G}
  \langle G \rangle
  = {2e^2N_1N_2 \over hN}
  + \delta G; \\ \fl
  \delta G =
  -
  {2e^2 N_1N_2\over hN(t_{\rm f} - t_{\rm i})  } 
  \int_{t_{\rm i}}^{t_{\rm f}} dt
  \int_0^{\infty} d\tau 
  \nonumber \\ \mbox{} \times
  \exp \left[-(N+4\alpha^2)\tau 
  -2 \int_0^{\tau} d\tau_1 (x(t-\tau+\tau_1) - x(t-\tau_1))^2 \right].
  \label{eq:deltaG} 
\end{eqnarray}
The correction term $\delta G$ of equation (\ref{eq:deltaG})
is the weak localization correction; it results from the
constructive interference of time-reversed trajectories. 
The presence of a time-dependent potential breaks time-reversal
symmetry and suppresses the weak localization correction.
Vavilov and Aleiner investigated the case
$x(t) = \delta x \cos(\omega t)$ of a harmonic time dependence
for the parameter $x$ in detail. In that case, the suppression
of weak localization increases with increasing frequencies and
saturates at a value
\begin{equation}
  \delta G = -{2 e^2 N_1 N_2 \over N^2 h} \times
  \left\{ \begin{array}{ll} \left[1-
  {2 (\delta x)^2 \over N + 4 \alpha^2 \vphantom{M^M_M}} 
  \right]
   &
  \mbox{if $(\delta x)^2 \ll N + 4 \alpha^2$}, \\
  \sqrt{N + 4 \alpha^2 
  \vphantom{M^M_M}
  \over 4 (\delta x)^2}
  & \mbox{if $(\delta x)^2 \gg N + 4 \alpha^2$,}
  \end{array} \right.
\end{equation}
for frequencies $\hbar\omega \gtrsim N \Delta$ \cite{VavilovAleiner}.
(Applicability of random matrix theory requires that $\omega 
\ll 1/\tau_{\rm erg}$, where $\tau_{\rm erg}$ is the time for
ergodic exploration of the quantum dot.)
If the fluctuations of the parameter $x$ are fast and random
on the scale $\hplanck/N \Delta$ of the delay time in the dot, 
they may be considered Gaussian white 
noise,
\begin{equation}
  \langle x(t) x(t') \rangle = {1 \over 4} \gamma \delta(t-t').
\end{equation}
In that case,
the exponent in equation (\ref{eq:deltaG}) can be averaged
separately, and one finds the result
\begin{equation}
  \delta G = -{2 e^2 N_1 N_2 \over h N (N+ 4 \alpha^2 + \gamma)}.
\end{equation}
The same suppression of weak localization was obtained previously 
to describe the decohering effect of the coupling to an external
bath \cite{BarangerMello,AleinerLarkin,BrouwerBeenakker97}. Note
that the strong-perturbation asymptote for white noise is different
from the strong-perturbation asymptote for fast harmonic variations
of the dot's shape. The cause for this difference is the 
existence of small time windows in which 
time-reversal symmetry is not violated 
near times $t$ with $\cos(\omega t) = \pm 1$ for harmonic
variations $\propto \cos(\omega t)$, while for 
a random time dependence of $x(t)$ no such special times around
which time-reversal symmetry is preserved exist 
\cite{transport,transport1b}. 

Similarly the variance of the conductance can also
be expressed in terms of the correlator (\ref{eq:W1t}).
(As in the time-independent case, the non-Gaussian 
correlator (\ref{eq:W2t}) does not contribute to the variance of
the conductance.) 
We refer to reference \onlinecite{VA2} for the
detailed expression for $\mbox{var}\, G$ and an analysis of
the effect of a harmonic time dependence of the shape 
function $x(t)$. Conductance fluctuations for the case
when $x(t)$ is a sum of two harmonics with
different frequencies were considered by
Kravtsov and Wang \cite{transport}.
\begin{figure}
\epsfxsize=0.7\hsize
\hspace{0.15\hsize}
\epsffile{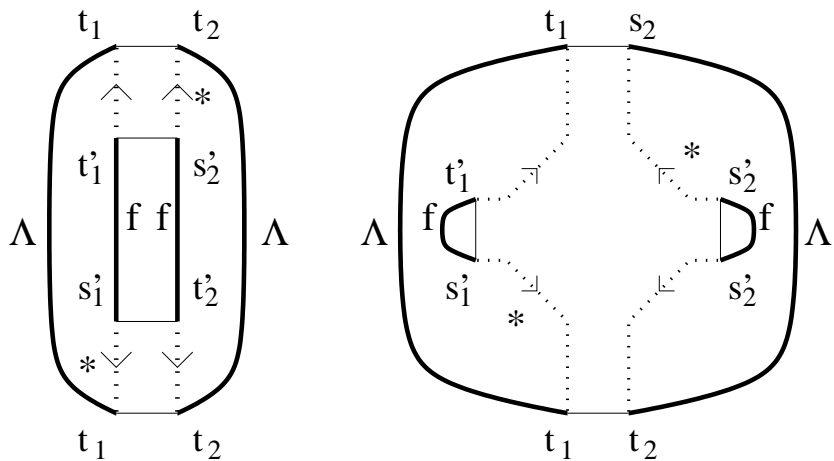}

\caption{\label{fig:pumpdiag}
Diagrams representing the two contributions to the variance
of the pumped current. Following the notation of reference
\onlinecite{unitary}, dotted lines correspond to the scattering
matrix ${\cal S}$, thick solid lines correspond to the fixed
matrices $\Lambda$ and $f$, and thin solid lines correspond
to the Kronecker delta's in the average over the ensemble of
scattering matrices.}
\end{figure}

{\em Pumped current.} To first order in the pumping frequency
$\omega$, the current in the absence of a source-drain voltage
is nonzero only if two or more parameters $x_j$
that determine the dot's shape are varied independently. Even
then, the ensemble average of the pumped current is zero, and the
first nonzero moment is $\langle I^2 \rangle$. The ensemble
average $\langle I^2 \rangle$ was calculated in reference \onlinecite
{B_p} for small pumping amplitudes $x_1(t) = \delta x_1 \sin(\omega t)$,
$x_2 = \delta x_2 \sin(\omega t + \phi)$,
\begin{equation}
  \langle I^2 \rangle^{1/2} = {e \omega \delta x_1 \delta x_2
  \over 2 \pi N} \sin \phi,
\end{equation}
independent of the presence or absence of a magnetic field.
The case of pumping amplitudes of arbitrary strength was 
considered in reference \onlinecite{SAA}.

Beyond the
adiabatic regime, one time-dependent parameter is sufficient to
generate a finite current through the dot
\cite{FalkoKhmelnitskii,spivak}. The second moment $\langle I^2
\rangle$ in that most general case was first calculated in reference
\onlinecite{VAA}, using the random Hamiltonian approach. The second 
moment, which involves an average
over four scattering matrix elements, can also be obtained in
the random scattering matrix approach, using
equations (\ref{eq:W2D})--(\ref{eq:F}) of the
previous section. We then find that there are two
contributions to $\langle I_{\rm pump}^2 \rangle$: one
contribution with two Gaussian contractions of scattering matrices
(giving a factor $W_1^2$) and one contribution which involves a
correlator of four scattering matrices (giving a factor $W_2$).
Diagrams representing these two contributions are shown in figure
\ref{fig:pumpdiag}. Adding both contributions, we find
\begin{eqnarray} \fl
  \langle I_{\rm pump}^2 \rangle &=&
  {8 e^2 N_1 N_2 \over N (t_{\rm f} - t_{\rm i})^2}
  \int_{t_{\rm i}}^{t_{\rm f}}
  dt dt' \int_0^{\infty} d\tau d\xi d\xi' \int_{-\tau}^{\tau} d\tau'
  f_{\rm eq}(2\tau') f_{\rm eq}(-2 \tau')
  \nonumber \\ \fl && \mbox{} \times
  {\cal D}(t,t-\tau-\tau';t',t'-\tau-\tau')
  {\cal D}(t',t'-\tau+\tau';t,t-\tau+\tau')
  \nonumber \\ \fl && \mbox{} \times
  {\cal
  D}(t-\tau-\tau',t-\tau-\tau'-\xi;t-\tau+\tau',t-\tau+\tau'-\xi)
  \nonumber \\ \fl && \mbox{} \times
  {\cal
  D}(t'-\tau+\tau',t'-\tau+\tau'-\xi';t'-\tau-\tau',t'-\tau-\tau'-\xi')
  \nonumber \\ \fl && \mbox{} \times
  \left[(\delta(\xi) - N)(\delta(\xi') -N) 
  + 4 \sum_{m} (x_m(t'-\tau+\tau') - x_m(t'-\tau-\tau'))
  \right. \nonumber \\ \fl && \left. \ \ \ \ \mbox{} \times
  (x_m(t-\tau+\tau') - x_m(t-\tau-\tau'))
  \vphantom{\sum_{m}} 
  \right],
\end{eqnarray}
independent of the value of the magnetic field.
Using equation (\ref{eq:Dmindiff}) of appendix \appc\
to express the delta functions
in terms of the functions $x_m(t)$ and a total
derivative of ${\cal D}$, performing partial integrations,
and shifting $t \to t-\tau$, $t'
\to t'-\tau$, this can be rewritten as
\begin{eqnarray}
  \fl
  \langle I_{\rm pump}^2 \rangle =
  {32 e^2 N_1 N_2 \over N (t_{\rm f} - t_{\rm i})^2}
  \int_0^{\infty} d\tau d\xi d\xi' \int_{-\tau}^{\tau} d\tau'
  \int_{t_{\rm i}-\tau}^{t_{\rm f}-\tau} dt dt'
  f_{\rm eq}(2\tau') f_{\rm eq}(-2 \tau')
  \nonumber \\ \mbox{} \times
  {\cal D}(t+\tau,t-\tau';t'+\tau,t'-\tau')
  {\cal D}(t-\tau',t-\tau'-\xi;t+\tau',t+\tau'-\xi)
  \nonumber \\ \mbox{} \times
  {\cal D}(t'+\tau,t'+\tau';t+\tau,t+\tau')
  {\cal D}(t'+\tau',t'+\tau'-\xi';t'-\tau',t'-\tau'-\xi')
  \nonumber \\ \mbox{} \times
  \left[\sum_{m,n}
  \left( \sum_{\pm} \pm x_m(t \pm \tau' - \xi) \right)^2 
  \left( \sum_{\pm} \pm x_n(t \pm \tau' - \xi') \right)^2 
  \right. \nonumber \\ \left. \ \ \mbox{}
  +
  \sum_{m} 
  \left( \sum_{\pm} \pm x_m(t'-\tau\pm \tau') \right)
  \left( \sum_{\pm} \pm x_m(t -\tau\pm \tau') \right)
  \right].
  \label{eq:pumpres}
\end{eqnarray}
This expression agrees with the result found by Vavilov, Ambegaokar,
and Aleiner \cite{VAA}. We refer
to reference \onlinecite{VAA} for a detailed analysis of equation
(\ref{eq:pumpres}) for the limiting
cases of adiabatic pumping and high-frequency pumping with one
and two time-dependent parameters.

{\em Noise.}
The current noise is defined as the variance of the charge transmitted
through the quantum dot
in the time interval $t_{\rm i} < t < t_{\rm f}$
\begin{equation}
  S = {1 \over t_{\rm f} - t_{\rm i}}
  \int dt dt' \left( \overline{{\bf I}(t) {\bf I}(t')}
  - \overline{{\bf I}(t)}\, \overline{{\bf I}(t')} \right).
\end{equation}
As in equation (\ref{eq:aaverage}),
$\overline{\cdots}$ denotes a quantummechanical or thermal
average, not an ensemble average. Performing the quantummechanical
and thermal average over the incoming states \cite{scatt1,scatt2,scatt3},
the noise power $S$ can be calculated as
\begin{eqnarray}
  \label{eq:noise}
  \fl
  S = {2 e^2 \over t_{\rm f} - t_{\rm i}}
  \int_{t_{\rm i}}^{t_{\rm f}} dt dt'\int dt_1 dt_2 dt'_1 dt'_2
  \tr\left[
  \left( {\cal S}^{\dagger}(t_1,t) \Lambda {\cal S}(t,t_2)
  - \delta(t_1-t) \Lambda \delta(t-t_2) \right)
  \right. \nonumber \\ \left. \mbox{} \times
  \tilde f(t_1'-t_2)
  \left( {\cal S}^{\dagger}(t_1',t') \Lambda {\cal S}(t',t_2')
  - \delta(t_1'-t') \Lambda \delta(t'-t_2') \right)
  f(t_1-t_2') \right].
  \nonumber
\end{eqnarray}
(A factor two has been added to account for spin degeneracy.) 
In the absence of a time-dependent potential, Equation (\ref{eq:noise}) 
represents the sum of Nyquist noise and shot noise \cite{BlanterBuettiker}.
With time-dependence, it contains an extra contribution to the noise
that is caused by the time dependence of the potential in the quantum
dot \cite{AK,L,AM,MM,buttikernew,PVB}.

Averaging equation (\ref{eq:noise}) for an ensemble of chaotic quantum
dots, we find 
\begin{eqnarray} \fl
 \langle S \rangle = S^N+S^S+S^P,\nonumber \\ \fl
 S^N =  2 k T h \langle G \rangle, \nonumber \\ \fl
 S^S =  e V h \langle G \rangle
  {N_1 N_2 \over 2 \pi N^2}
 \left(\coth\frac{e V}{2kT}-\frac{2k T}{e V}\right),  
 \nonumber\\ \fl
 S^P = {e^2 N_1 N_2 (kT/\hbar)^2 \over 2 N (t_{\rm f} - t_{\rm i})}
 \int_{t_{\rm i}}^{t_{\rm f}}dt dt'
 \left( {1 \over N^2} - 
 \left[\int_0^\infty {\cal D}(t,t-\xi;t',t'-\xi)d\xi
 \right]^2\right)\nonumber \\ \mbox{}
 \times
  \frac{N^2-2({N_1^2+N_2^2})\sin^2 [e V(t-t')/2 \hbar]}
  {\sinh^2[\pi k T (t-t')/\hbar]},
  \label{eq:noises}
\end{eqnarray}
where $\langle G \rangle$ is the average (time-dependent)
conductance, see equation (\ref{eq:G}), and $V = (\mu_1
-\mu_2)/e$ the bias voltage.
The above ensemble averages for the Nyquist noise and shot
noise 
are the same as the noise power found in the absence of a
time-dependent potential \cite{Nazarov}, up to an eventual
weak localization correction. The extra
noise generated by the time-dependence of the dot shape is
fully described by the term $S^P$.
In the adiabatic regime $\hbar \omega \ll N \Delta$, the 
pumping noise can be written as
\begin{eqnarray}\label{eq:noisepower} \fl
  S^P = \frac{e^2 N_1 N_2 (kT/\hbar)^2}{2 N (t_{\rm f}-t_{\rm i})}
  \int_{t_{\rm i}}^{t_{\rm f}}dt'dt
\left[{1 \over N^2} - \left(\frac{1}{N+2 
  \sum_{m} (x_m(t) - x_m(t'))^2}\right)^2\right]\nonumber \\
  \mbox{} \times
  \frac{N^2-2({N_1^2+N_2^2})\sin^2 [e V(t-t')/2 \hbar]}
  {\sinh^2[\pi k T (t-t')/\hbar]},
\end{eqnarray}
In the absence of a bias voltage, $eV = 0$,
Equation (\ref{eq:noisepower}) has been analyzed in detail 
by Vavilov and the authors in reference \onlinecite{PVB}.
For one time-dependent parameter $x(t) = 
\delta x \cos(\omega t)$ it is found that
\begin{eqnarray} \fl
  S^P = {\omega e^2 N_1 N_2 \over \pi^2 N^2} 
  \left\{ \begin{array}{ll}
  2 \pi (\delta x)^2 \left(
  \coth{\hbar \omega \over 2 k T} 
  - {2 k T \over
  \hbar \omega} \right)_{\vphantom{M}}^{\vphantom{M}},
  & \mbox{if $ (\delta x)^2 \ll N \max(1,k^2 T^2/ \hbar^2 \omega^2)$,} \\
  \vphantom{\left( M^M_M \right)}
  3 |\delta x| N^{1/2}, & 
  \mbox{if $ (\delta x)^2 \gg N \max(1, k^2 T^2/ \hbar^2 \omega^2)$.}
  \end{array} \right.
\end{eqnarray}
An applied bias voltage has an effect on the pumping noise
$S^P$ only if $eV \gtrsim \max(\hbar \omega,k T,
\hbar \omega |\delta x|/N^{1/2})$. And even then, the effect
of the applied bias is limited to a reduction of $S^P$ by a 
numerical factor $2 N_1 N_2/N^2$. In this respect, the effect
of an external bias on the pumping noise is much weaker than
that of temperature, which tends to suppress $S^P$ as soon
as $k T \gtrsim \hbar \omega \max(1,|\delta x|/N^{1/2})$
\cite{PVB}. 

\section{Conclusion}

In summary, in this paper we have extended the scattering
approach of the random-matrix theory of quantum transport to
the case of scattering from a chaotic quantum dot with a time-dependent
potential. We addressed the limit that the
number of channels $N$ coupling the dot to the electron
reservoirs is large. In this limit, the elements of the scattering
matrix have a distribution that is almost Gaussian, with non-Gaussian
corrections that are small as $N$ becomes large. We calculated the
second moment, which defines the Gaussian part of the
distribution, and the
fourth cumulant, which characterizes the leading non-Gaussian
corrections.

The advantage of the scattering matrix approach is that, once the
scattering matrix distribution is calculated, the computation of
transport properties is a matter of mere quadrature. As an
example, we calculated the conductance of a quantum dot with a
time-dependent potential or the current pumped through the
dot in the absence of an external bias, and found agreement
with previous calculations of Vavilov {\em et al.} that were
based on the Hamiltonian approach \cite{VavilovAleiner,VA2,VAA}. 
The results derived here were used for the calculation of the 
current noise generated by the time-dependence of the potential 
in the quantum dot by Vavilov and the authors \cite{PVB}. 
The current noise in the presence
of both a time-dependent potential in the dot and a bias
voltage was studied here.

Whereas the first four moments of the scattering matrix
distribution that we calculated here are sufficient for the
calculation of most transport properties --- most transport
properties are quadratic or quartic in the scattering matrix ---,
we need to point out that there are observables that cannot
be calculated with the results presented here.
First, in the presence of one or more
superconducting contacts, (averaged) transport properties may
still depend on higher cumulants of the distribution, despite the
fact that these are small by additional factors of $1/N$
\cite{BeenakkerReview}.
Second, the results presented here fail to quantitatively
describe transport properties for very small $N$, which can
have strongly non-Gaussian distributions. Further research
in these directions is necessary.

\ack

We would like to thank Maxim Vavilov for important contributions.
This work was supported by the Cornell Center for Materials
Research under NSF grant no.\ DMR-0079992, by the NSF under
grant no.\ 0086509, and by the Packard Foundation.

\appendix

\section{Nonideal contacts}
\label{app:b}

Nonideal contacts are characterized by channels that have a transmission
coefficient $\Gamma_j$ smaller than unity, $j=1,\ldots,N$.
The imperfect transmission
of the contacts is characterized by an $N \times N$
reflection matrix $r_c(t,t')$, for which we take the simple form
\begin{equation}
  r_c(t,t') = (1-\Gamma)^{1/2} \delta(t-t'),
  \label{eq:rc}
\end{equation}
where $\Gamma$ is an $N \times N$ diagonal matrix containing the
transmission coefficients $\Gamma_j$ on the diagonal.
The direct backscattering from the contacts is fast
compared to the scattering that involves ergodic exploration of the
dot, hence the delta function
$\delta(t-t')$ in equation (\ref{eq:rc}). In order to describe
time dependent scattering with nonideal leads, we use a
modification of the stub model of equation (\ref{eq:stub})
\cite{FriedmanMello,Brouwer1995},
\begin{eqnarray}
  {\cal S} &=& r_c +
  \Gamma^{1/2} S^{\rm fl} \Gamma^{1/2},\ \
  S^{\rm fl} =
  P{\cal U}(1-R {\cal U})^{-1}P^\dagger
  , \label{eq:rcstub}
  \\
  R &=& Q^{\dagger} \rme^{-2 \pi \rmi H/M \Delta} Q - P^{\dagger} r_c P.
  \label{eq:RHrc}
\end{eqnarray}
The first term in equation (\ref{eq:rcstub}) takes into account the
direct backscattering at the contact for electrons coming in
from the reservoirs, whereas the extra term
in equation (\ref{eq:RHrc}) describes backscattering at
the contact for electrons coming from the dot. The additional
factors $\Gamma^{1/2}$ in the second term of equation (\ref{eq:rcstub})
account for the decreased
transmission probability for entering or exiting the quantum dot.
With the inclusion of reflection in the contacts as in equation
(\ref{eq:rcstub}), the scattering matrix approach for 
time-independent scattering was proven to be fully equivalent
to the Hamiltonian approach with arbitrary coupling to the
leads \cite{Brouwer1995}. The corresponding distribution of
the scattering matrix ${\cal S}$ for time-independent 
scattering is known as the Poisson kernel \cite{MPS}.

Like in the case of ideal leads, the distribution of the
elements of the
scattering matrix ${\cal S}$ for a quantum dot with
nonideal leads is almost Gaussian, with
non-Gaussian corrections that are small if $N \gg 1$. The main
difference with the case of an ideal contact is that,
as a result of the direct reflection from the contact, the average
of ${\cal S}$ is nonzero for a nonideal contact. The fluctuations
of ${\cal S}$ around the average are described by ${\cal S}^{\rm fl}$,
cf.\ equation (\ref{eq:rcstub}).
In order to find the distribution of ${\cal S}^{\rm fl}$, we
note that the expression (\ref{eq:rcstub})
for ${\cal S}^{\rm fl}$
is formally equivalent to the stub
model equation (\ref{eq:stub}) used to describe time-dependent
scattering from a quantum dot with ideal contacts. Hence we conclude
that the moments of ${\cal S}^{\rm fl}$ can be obtained directly
from the results for the case of ideal contacts, see section
\ref{sec:2} and appendix
\appc, provided we substitute equation (\ref{eq:RHrc}) for
the matrix $R$. This amounts to the replacement 
${\cal S} \to {\cal S}^{\rm fl}$ in the final results 
(\ref{eq:W2D})--(\ref{eq:F}), $N \to \sum_{j} \Gamma_j$ in
equation (\ref{eq:Dresult}), and $N \to \sum_{j} \Gamma_j^2$
in equation (\ref{eq:F}).

\section{Correlators for time-independent scattering}
\label{app:a}

The scattering matrix correlators for time-independent
scattering serve as input for the calculation of the
correlators for time-dependent scattering. They can be
calculated using the Hamiltonian approach (see references
\onlinecite{VWZ,Efetov95,Frahm95}), or, alternatively,
in the scattering matrix approach,
using a time-independent version of the ``stub model''
of section \ref{sec:2}. Following the latter method, the scattering
matrix ${\cal S}$ is written as \cite{BB}
\begin{eqnarray}\label{eq:stub2}
{\cal S}(\varepsilon)=P{\cal U}(1-R {\cal U})^{-1}P^\dagger,\ \
  R = Q^{\dagger} \rme^{2 \pi \rmi \varepsilon/M \Delta} Q.
  \label{eq:RH2}
\end{eqnarray}
Here the matrices $P$ and $Q$ are as in equation (\ref{eq:stub}),
whereas ${\cal U}$ is an $M \times M$ unitary matrix taken from
the circular orthogonal ensemble or circular unitary ensemble
of random matrix theory, depending on the presence or absence
of time-reversal symmetry. The picture underlying
equation (\ref{eq:RH2}) is that a stub with $M-N$ scattering
channels is attached to the chaotic
quantum dot as in figure \ref{fig:stub}, 
such that the dwell time in the stub is much larger than
the dwell time in the dot, but much smaller than the total dwell
time in the combined dot-stub system.
The first condition implies that the
$M \times M$ scattering matrix of the
chaotic dot (without stub)
may be taken energy independent, and distributed
according to the appropriate circular ensemble from random
matrix theory. The total scattering matrix ${\cal S}$ then
acquires its energy dependence through the energy dependence
of the $(M-N)\times(M-N)$ reflection matrix $R$ of the stub.
The second condition, which requires $M \gg N$, ensures
that the dot plus stub system is explored ergodically before
an electron escapes into the lead, so that 
the spatial separation of the energy dependence (stub)
and chaotic scattering (dot))
does not affect the correlators
of the scattering matrix ${\cal S}$.\footnote{Note that this
version of the ``stub model'' is different from that used in the
main text. In time representation, the matrix ${\cal U}$ of
equation (\ref{eq:RH2}) is proportional to a delta function
$\delta(t-t')$, whereas the matrix $R$ involves a time delay
with time $t-t'=\hplanck/M \Delta$. For the model of section
\ref{sec:2} of the main text, the time delay is described by
${\cal U}$, whereas scattering from the stub is instantaneous.
Both versions of the ``stub model'' are equivalent to the 
Hamiltonian approach.
Which one to use is a matter of convenience.}

Using the diagrammatic technique of reference \onlinecite{unitary}
to average over the random unitary matrix ${\cal U}$,
we find that the second moment $W_1$ is given by
\begin{eqnarray} \fl
  W_1^{ij;kl}(\varepsilon;\varepsilon')
  &=& {1 \over M - \tr R(\varepsilon) R^{\dagger}(\varepsilon')}
  \times \left\{ \begin{array}{ll}
  ( \delta_{ik} \delta_{jl} + \delta_{il} \delta_{jk})
  & \mbox{with TRS}, \\
  \delta_{ik} \delta_{jl} &
  \mbox{without TRS}.
  \end{array} \right.
  \label{eq:Sdistrparamapp} \label{eq:cum1app}
\end{eqnarray}
Substitution of equation (\ref{eq:RH2}) for $R$ gives equation
(\ref{eq:cum1}) of section \ref{sec:2}. Note that equation
(\ref{eq:cum1}) is valid in the semiclassical limit of
large $N$ only. 
Within the diagrammatic technique this follows from the
observation that for large $N$ the only contributions to $W_1$ are
the ``ladder'' and ``maximally crossed'' diagrams, whereas 
for small $N$ more contributions exist and a non-perturbative
calculation is needed to calculate the scattering
matrix correlator \cite{unitary}.
The correlator $W_1(\varepsilon,\varepsilon')$ was calculated
by Verbaarschot {\em et al.} in
reference \onlinecite{VWZ} for arbitrary $N$ using the Hamiltonian
approach and the supersymmetry technique.

For the cumulant $W_2$ we find in the absence of time-reversal
symmetry
\begin{eqnarray}\fl
  W_2^{i_1j_1,i_2j_2;k_1l_1,k_2l_2}
  (\varepsilon_1,\varepsilon_2;\varepsilon_1',\varepsilon_2')
  = -
  \left(
  \delta_{i_1k_1} \delta_{j_1 l_2} \delta_{i_2 k_2} \delta_{j_2 l_1}
  +
  \delta_{i_1k_1} \delta_{j_1 l_2} \delta_{i_2 k_2} \delta_{j_2 l_1}
  \right)
  \nonumber \\ \mbox{} \times
  [M - \tr R(\varepsilon_1) R(\varepsilon_2)
  R^{\dagger}(\varepsilon_1') R^{\dagger}(\varepsilon_2')]
  \nonumber \\ \mbox{} \times
  [M - \tr R(\varepsilon_1) R(\varepsilon_1')]^{-1}
  [M - \tr R(\varepsilon_2) R(\varepsilon_1')]^{-1}
  \nonumber \\ \mbox{} \times
  [M - \tr R(\varepsilon_1) R(\varepsilon_2')]^{-1}
  [M - \tr R(\varepsilon_2) R(\varepsilon_2')]^{-1}.
   \label{eq:W2app}
\end{eqnarray}
In the presence of time-reversal symmetry, fourteen terms corresponding
to the permutations
$i_2 \leftrightarrow j_2$, $k_1 \leftrightarrow l_1$, and
$k_2 \leftrightarrow l_2$ have to be added. Equation (\ref{eq:cum2})
is recovered upon substitution of equation (\ref{eq:RH2}) for $R$.

\section{Correlators for
time-dependent scattering} \label{sec:3}

In this appendix we present the derivations of equations
(\ref{eq:W2D})--(\ref{eq:F}). 

We first calculate the second moment $W_1$ of the scattering
matrix distribution,
equations (\ref{eq:W2D}) and (\ref{eq:Dresult}).
To find $W_1$ we use equation (\ref{eq:stub}) to
expand ${\cal S}$ in powers of
${\cal U}$ and $R$ and then average
over ${\cal U}$. In the limit of large $M$ and large $N$, that
average can be done using the cumulants (\ref{eq:cum1t}) and
(\ref{eq:cum2t}) and the diagrammatic rules
of reference \onlinecite{unitary}. This calculation is similar
to the standard diagrammatic perturbation theory: the
matrices ${\cal U}$, ${\cal U}^{\dagger}$, and $R(t)$
play the role of the unperturbed retarded and
advanced Green functions and the random potential,
respectively.

\begin{figure}
\epsfxsize=0.65\hsize
\hspace{0.25\hsize}
\epsffile{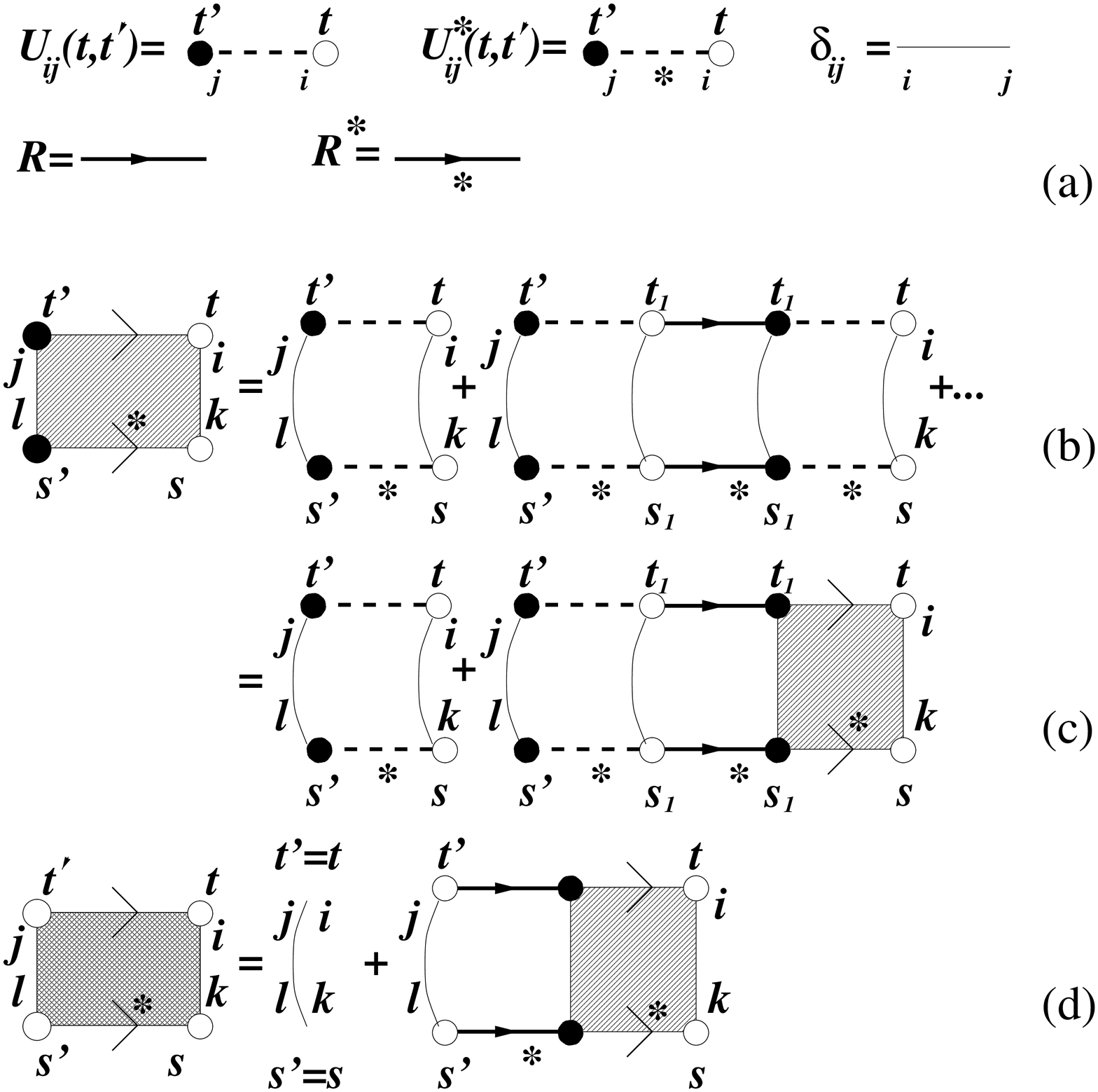}
\vglue +0.2cm
\caption{\label{fig:1}
(a) Notations, following reference \protect\onlinecite{unitary}.
(b) Calculation of the kernel $D(t,t';s,s')$.
(c) Diagrammatic representation of the
  Dyson equation (\ref{eq:Dtau}) for $D(t,t-\tau;s,s-\tau)$, with
$t' = t-\tau$, $t_1 = t - \tau_1$, $s' = s-\tau$, and $s_1 = s-\tau_1$.
(d) Diagrammatic representation of the differential
equation (\protect\ref{eq:Dmindiff})
for $D(t,t-\tau;s,s-\tau)$. Left hand side: $(M+\partial_{\tau})
D(t,t-\tau;s,s-\tau)$; Right hand side:
$\delta(\tau) + \tr R(t-\tau) R^{\dagger}(s-\tau)
D(t,t-\tau;s,s-\tau)$.}
\end{figure}

Performing the average over ${\cal U}$ this way, we find that,
to leading order in $M^{-1}$ and $N^{-1}$, the cumulant
$W_1$ is dominated by two leading contributions: the
``ladder diagram'' of figure \ref{fig:1} and the
``maximally crossed diagram'' of figure \ref{fig:2}.
Since every factor in these two diagrams involves equal time differences for
${\cal S}$ and ${\cal S}^*$,
we conclude that this contribution to $W_1$ is nonzero only if $t-t'=
s-s'$, cf.\ equation (\ref{eq:cum1t}). Further, we conclude that the
ladder diagram gives a nonzero contribution only if $i=k$ and $j=l$,
while the maximally crossed diagram contributes when $i=l$ and $j=k$.

We first consider the contribution of the ladder diagram, which we
write as
\begin{equation}
  \delta_{ik} \delta_{jl}
  \delta(t-t'-s+s')
  \Dbare(t,t';s,s'),
  \label{eq:Dplusdef}
\end{equation}
where the kernel $\Dbare$ is the equivalent of the ``diffuson''
from standard diagrammatic perturbation theory. Note that, in view
of the delta function in equation (\ref{eq:Dplusdef}), the kernel
$\Dbare$ depends on three arguments, not on four. For notational
convenience, we prefer, however, to continue to use the two
initial times $t'$ and $s'$ and the two final times $t$ and $s$
to denote the time-arguments of $\Dbare$.

Considering the ladder diagrams to all orders, the diffuson $\Dbare$
is found to obey the Dyson equation
\begin{eqnarray} \fl
  \Dbare(t,t-\tau;s,s-\tau)
  = \theta(\tau) \rme^{-M\tau} +
  \theta(\tau) \int_{0}^{\tau} d \tau_1
  \Dbare(t,t-\tau_1;s,s-\tau_1)\,
  \nonumber \\ \mbox{} \times
  \tr R(t-\tau_1) R^\dagger(s-\tau_1)
  \rme^{-M(\tau-\tau_1)}.
  \label{eq:Dtau}
\end{eqnarray}
The solution of equation (\ref{eq:Dtau}) is
\begin{eqnarray} \fl
  \Dbare(t'+\tau,t';s'+\tau,s') &=&
  \theta(\tau)
  \exp \left[-\int_{0}^{\tau} d \tau_1
  \left(M-\tr R(t'+\tau_1)R^\dagger(s'+\tau_1) \right) \right],
  \label{eq:Dsol}
\end{eqnarray}
where we used that $D=0$ if $\tau < 0$. 
Substitution of $R = \exp(2\pi H/\Delta)$ reproduces the first
term in the result (\ref{eq:W2D}). For future use, 
we note that the function $\Dbare$ of equation (\ref{eq:Dsol})
obeys the differential equations
\begin{eqnarray}
  \fl
  {{\partial \over \partial \tau}
  \Dbare(t,t-\tau;s,s-\tau)}
  = \delta(\tau)
  -\left[M - \tr R(t-\tau) R^\dagger(s-\tau)\right]
  \Dbare(t,t-\tau;s,s-\tau), \label{eq:Dmindiff}
  \\ \fl
  {{\partial \over \partial \tau}
  \Dbare(t'+\tau,t';s'+\tau,s')} 
  = \delta(\tau)
  -\left[M - \tr R(t'+\tau) R^\dagger(s'+\tau)\right]
  \Dbare(t'+\tau,t';s'+\tau,s').
  \nonumber \\
  \label{eq:Ddiff}
\end{eqnarray}

\begin{figure}
\epsfxsize=0.65\hsize
\hspace{0.2\hsize}
\epsffile{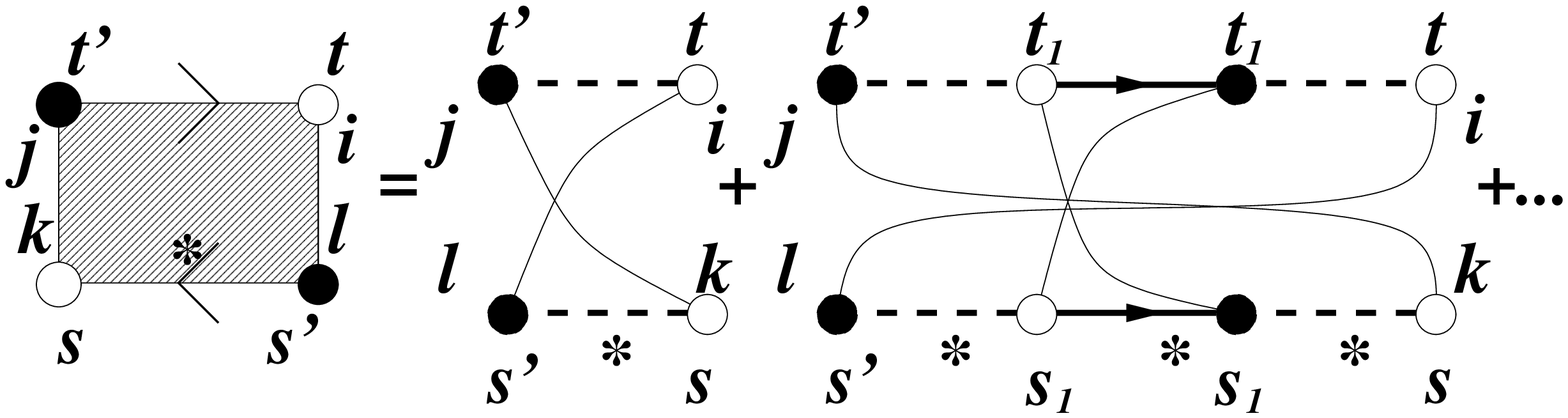}
\vglue +0.2cm
\caption{\label{fig:2}
Diagram representing the
kernel $C(t,t';s,s')$ of equation (\protect\ref{eq:cooperon}).}
\end{figure}

Calculation of the contribution of the maximally crossed diagram
proceeds in an analogous way. This contribution reads
\begin{eqnarray}
  \delta(t-t'-s+s')
  \delta_{il} \delta_{jk} \Cbare(t,t';s,s'),
  \label{eq:CD}
\end{eqnarray}
where the analogue of the Cooperon is given by
\begin{equation} \fl
  \Cbare(t'+\tau,t';s'+\tau,s') =
  \theta(\tau)
  \exp \left[
  {-\int_{0}^{\tau} d\tau_1
  \left(M-\tr R(t'+\tau_1)R^\dagger(s'+\tau-\tau_1) \right)} \right].
  \label{eq:cooperon}
\end{equation}
Substitution of $R = \exp(2 \pi H/\Delta)$ gives the second term
of equation (\ref{eq:W2D}).

We now turn to the four scattering-matrix correlator
(\ref{eq:W2tdef}), which is the equivalent of the Hikami box
in standard diagrammatic perturbation theory. 
We first calculate the first term of equation (\ref{eq:W2t}).
It is represented diagrammatically in
figure \ref{4corr}. There are two contributions:
One contribution involving Gaussian
contractions with the cumulant (\ref{eq:cum1t}) only, which
is depicted as the first term on the r.h.s.\ of 
figure \ref{4corr}, and one contribution that involves the
non-Gaussian contraction of equation (\ref{eq:cum2t}) once and
otherwise Gaussian contractions, see the second term on the
r.h.s.\ of
figure \ref{4corr}. Adding those two contributions, the
function ${\cal F}$ is found to be,
\begin{eqnarray} \fl
  {\cal F}(t_1,t_1';t_2,t_2';s_1,s_1';s_2,s_2') =
  \int d\tau \Dbare(t_1'+\tau,t_1';s_1'+\tau,s_1')
  \nonumber \\  \fl \ \  \mbox{} \times
  \Dbare(t_2-s_1+s_1'+\tau,t_2';s_2-t_1+t_1'+\tau,s_2')
  \nonumber \\ \fl \ \  \mbox{} \times
  \Dbare(t_1,t_1'+\tau;s_2,s_2-t_1+t_1'+\tau)
  \Dbare(t_2,t_2-s_1+s_1'+\tau;s_1,s_1'+\tau)
  \nonumber \\ \fl \ \  \mbox{} \times
  \tr R(t_1'+\tau) R^{\dagger}(s_2-t_1+t_1'+\tau)
  R(t_2-s_1+s_1'+\tau) R^{\dagger}(s_1'+\tau)
  \nonumber \\ \fl \ \   \mbox{}
  -
  \int d\tau_1 d\tau_2 d\tau_3 d\tau_4
  (M+\partial_{\tau_1}) \Dbare(t_1'+\tau_1,t_1';s_1'+\tau_1,s_1')
  \nonumber \\  \fl \ \  \mbox{} \times
  (M+\partial_{\tau_3}) \Dbare(t_2'+\tau_3,t_2';s_2'+\tau_3,s_2')
  (M+\partial_{\tau_2}) \Dbare(t_1,t_1-\tau_2;s_2,s_2-\tau_2)
  \nonumber \\ \fl \ \  \mbox{} \times
  (M+\partial_{\tau_4}) \Dbare(t_2,t_2-\tau_4;s_1,s_1-\tau_4)
  \nonumber \\ \fl \ \  \mbox{} \times
  \theta(t_1-t_1'-\tau_1-\tau_2)
  \theta(t_2-t_2'-\tau_3-\tau_4)
  \theta(s_1-s_1'-\tau_1-\tau_4)
  \theta(s_2-s_2'-\tau_2-\tau_3)
  \nonumber \\ \fl \ \ \mbox{} \times
  {\cal F}^0(t_1-t_1'-\tau_1-\tau_2;
  t_2-t_2'-\tau_3-\tau_4;
  s_1-s_1'-\tau_1-\tau_4;
  s_2-s_2'-\tau_2-\tau_3).  \label{eq:Fintermed}
\end{eqnarray}
Here we used equation (\ref{eq:Ddiff}) to express the four legs of the
diagrams of figure \ref{4corr}b in terms of the diffuson $\Dbare$
and its derivative. The second term in equation (\ref{eq:Fintermed}) can
be simplified noting that the time integration is dominated
by all four arguments of ${\cal F}^0$ being of order $1/M$.
Using the smallness of these time arguments, the diffusons can
be expanded around $t_1-t_1'-\tau_1-\tau_2=
t_2-t_2'-\tau_3-\tau_4= s_1-s_1'-\tau_1-\tau_4= s_2-s_2'-\tau_2-\tau_3
= 0$ and three of four time integrations can be done. The result is
\begin{eqnarray} \fl
  {\cal F}(t_1,t_1';t_2,t_2';s_1,s_1';s_2,s_2') =
  \int d\tau \Gamma(t_1,t_1';t_2,t_2';s_1,s_1';s_2,s_2';\tau)
  \Dbare(t_1'+\tau,t_1';s_1'+\tau,s_1')
  \nonumber \\ \mbox{} \times
  \Dbare(t_1,t_1'+\tau;s_2,s_2-t_1+t_1'+\tau)
  \Dbare(t_2,t_2-s_1+s_1'+\tau;,s_1,s_1'+\tau)
  \nonumber \\ \mbox{} \times
  \Dbare(t_2-s_1+s_1'+\tau,t_2';s_2-t_1+t_1'+\tau,s_2'),
  \label{eq:Ff}
\end{eqnarray}
where we abbreviated
\begin{eqnarray} \fl
  \Gamma =
  M- \tr R(t_1'+\tau) R^{\dagger}(s_2-t_1+t_1'+\tau)
  - \tr R(t_2-s_1+s_1'+\tau) R^{\dagger}(s_1'+\tau)
  \nonumber \\ \mbox{} +
  \tr R(t_1'+\tau) R^{\dagger}(s_2-t_1+t_1'+\tau)
  R(t_2-s_1+s_1'+\tau) R^{\dagger}(s_1'+\tau)
  \nonumber \\ \mbox{}
  -\delta(t_1-t_1'-\tau) - \delta(s_1-s_1'-\tau).
  \label{eq:Gamma1}
\end{eqnarray}

\begin{figure}
\epsfxsize=0.84\hsize
\hspace{0.15\hsize}
\epsffile{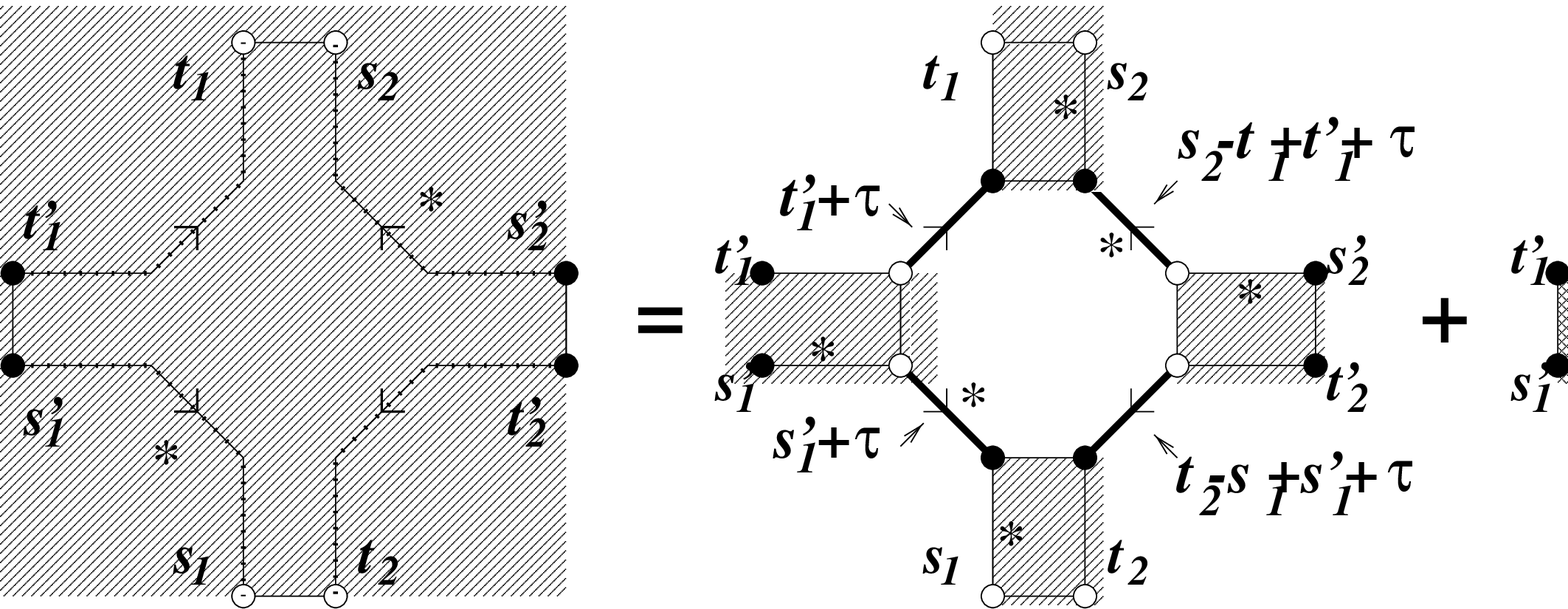}

\caption{\label{4corr}
Diagrams representing the two contributions to the correlator
${\cal F}(t_1,t_1';t_2,t_2';s_1,s_1';s_2,s_2')$ (l.h.s.). The
first diagram on the r.h.s.\
contains Gaussian contractions only; the shaded blocks
in denote the kernel $D$. The second diagram on the r.h.s.\
contains one
non-Gaussian contraction, which is in the center of the
diagram; the shaded blocks represent factors of the
form $(M+\partial_{\tau})D$,
see equation (\protect\ref{eq:Fintermed}) and
figure \protect\ref{fig:1}d.}
\end{figure}

\noindent
We used equations (\ref{eq:Ddiff}) and (\ref{eq:Dmindiff}) to
calculate time derivatives of the diffusons.
Alternatively, using a partial integration, the function ${\cal F}$
can be expressed by equation (\ref{eq:Ff}) with
\begin{eqnarray}
  \fl  \Gamma =
  M- \tr R(t_1'+\tau) R^{\dagger}(s_1'+\tau)
  - \tr R(t_2-s_1+s_1'+\tau) R^{\dagger}(s_2-t_1+t_1'+\tau)
  \nonumber \\ \mbox{} +
  \tr R(t_1'+\tau) R^{\dagger}(s_2-t_1+t_1'+\tau)
  R(t_2-s_1+s_1'+\tau) R^{\dagger}(s_1'+\tau)
  \nonumber \\ \mbox{}
  -\delta(\tau) - \delta(\tau-t_1+t_1'+s_2-s_2').
  \label{eq:Gamma2}
\end{eqnarray}
or with $\Gamma$ given by a convenient linear combination of equations
(\ref{eq:Gamma1}) and (\ref{eq:Gamma2}) with coefficients
 $C_1, C_2$ satisfying the condition $C_1+C_2=1$.

Finally, using equation (\ref{eq:RH}) for
$R$, the first term of equation (\ref{eq:W2t}) is obtained.
The other contributions to equation (\ref{eq:W2t}) can be found
after permutation of the channel indices and time-variables
as indicated in figure \ref{4corrdef}.

\end{document}